\documentclass[preprint,12pt]{elsarticle}

\usepackage{lineno,hyperref}
\usepackage[utf8]{inputenc} 
\usepackage[T1]{fontenc}    
\usepackage{hyperref}       
\usepackage{url}            
\usepackage{booktabs}       
\usepackage{amsfonts}       
\usepackage{nicefrac}       
\usepackage{microtype}      
\usepackage{lipsum}
\usepackage{amsmath}
\usepackage{graphicx}
\usepackage{subfigure}
\usepackage{tabularx}
\usepackage{array}
\usepackage{multirow}
\usepackage{caption}
\usepackage[title,toc,titletoc,header]{appendix}
\usepackage[T1]{fontenc}
\usepackage{lmodern}
\usepackage{algorithm, algorithmicx, algpseudocode}
\usepackage{xcolor,mathtools, comment}
\usepackage{natbib}
\modulolinenumbers[5]

\begin{document}

\begin{frontmatter}

\title{GGNB: Graph-Based Gaussian Naive Bayes Intrusion Detection System for CAN Bus}



\author[myaddress]{Riadul Islam\corref{mycorrespondingauthor}} 
\ead{riaduli@umbc.edu}
\author[myaddress]{Maloy K. Devnath}
\author[mythirdaddress]{Manar D. Samad}
\author[fourthaddress]{Syed Md Jaffrey Al Kadry}

\cortext[mycorrespondingauthor]{Riadul Islam}

\address[myaddress]{Department of Computer Science and Electrical Engineering, University of Maryland, Baltimore County, MD 21250, United States}
\address[mythirdaddress]{Department of Computer Science, Tennessee State University, TN 37209, United States}
\address[fourthaddress]{General Motors Corporation, Detroit, MI, United States}

\begin{abstract}
The national highway traffic safety administration (NHTSA) identified cybersecurity of the 
automobile systems are more critical than the security of other information systems.
Researchers already demonstrated remote attacks on critical vehicular electronic control units 
(ECUs) using controller area network (CAN). Besides, existing intrusion detection systems (IDSs) often propose to tackle a specific type of attack, 
which may leave a system vulnerable to numerous other types of attacks.
A generalizable IDS that can identify a wide range of attacks within the shortest possible 
time has more practical value than attack-specific IDSs, which is not a trivial task to accomplish.
In this paper we propose a novel {\textbf g}raph-based {\textbf G}aussian 
{\textbf n}aive {\textbf B}ayes (GGNB) intrusion detection algorithm by leveraging graph properties and PageRank-related features.
The GGNB on the real rawCAN data set~\cite{Lee:2017} yields 99.61\%, 99.83\%, 96.79\%, and 96.20\%  detection accuracy for denial of service (DoS), fuzzy, spoofing, replay, mixed attacks, respectively. Also, using  OpelAstra data set~\cite{Guillaume:2019}, the proposed methodology has 100\%, 99.85\%, 99.92\%, 100\%, 99.92\%, 97.75\% and 99.57\% detection accuracy considering DoS, diagnostic, fuzzing CAN ID, fuzzing payload, replay, suspension, and mixed attacks, respectively.  
The GGNB-based methodology requires about $239\times$ and $135\times$ lower training and tests times, respectively, compared to the SVM classifier used in the same application.
Using Xilinx Zybo Z7 field-programmable gate array (FPGA) board, the proposed GGNB requires $5.7 \times$, $5.9 \times$, $5.1 \times$, and $3.6 \times$ fewer slices, LUTs, flip-flops, and DSP units, respectively, than conventional NN architecture.
\end{abstract}


\end{frontmatter}


\section{Introduction}
\label{sec:intro}

The application of vehicular electronic control units (ECUs) depends on intra- and inter-vehicular communication networks' performance. These communications are vehicle-to-vehicle, vehicle-to-infrastructure, and vehicle-to-cloud. Any cyber or physical attack on these communication networks has severe consequences, including private data leakage, financial loss, and jeopardizing human life. As a result, it is imperative to secure these networks of modern vehicular systems. 

\begin{figure}[t]
	\begin{center}
		\vspace{-0.15cm} 
		\includegraphics[width = 0.85\textwidth]{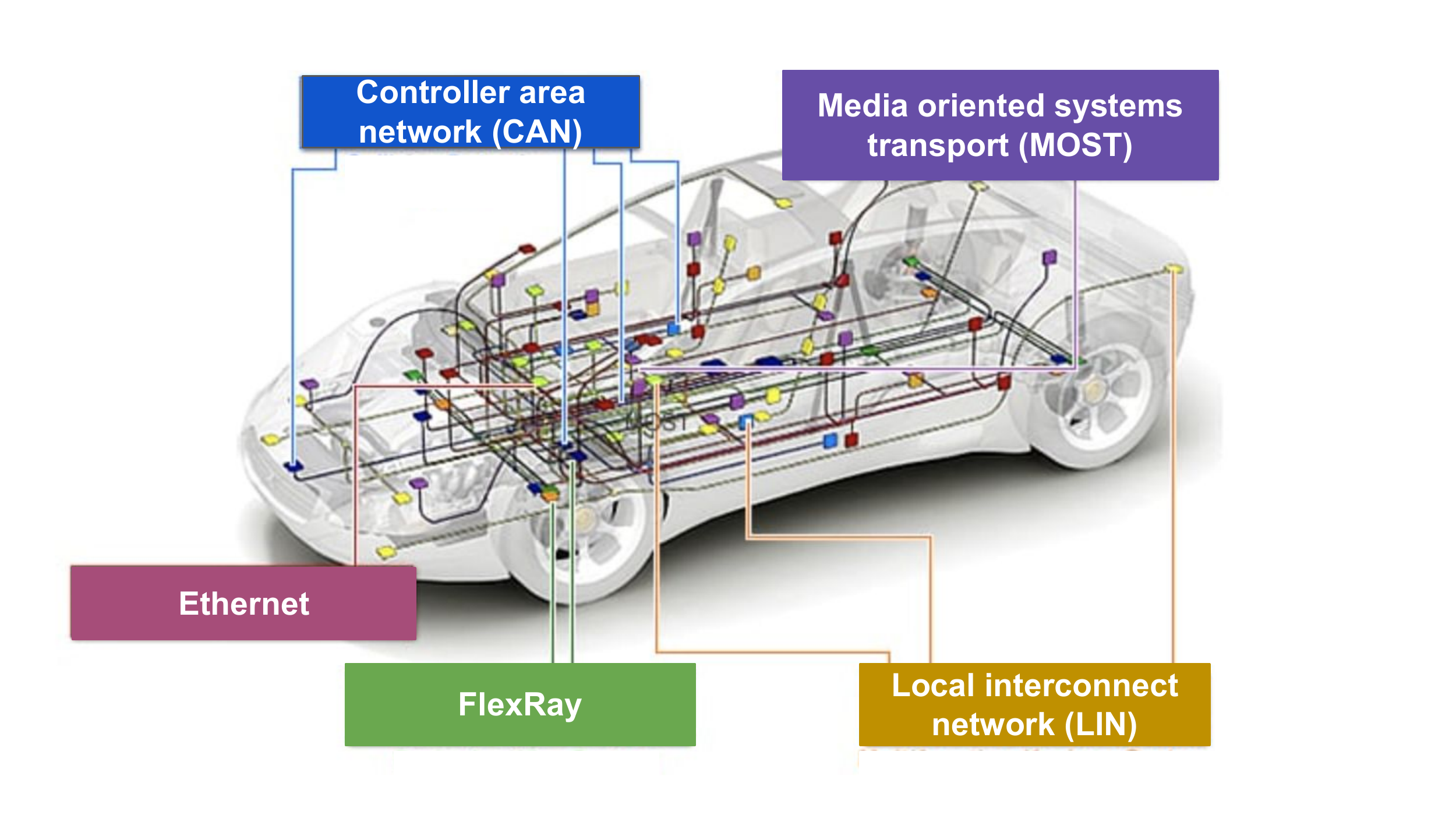}
		\vspace{-0.5cm} 
		\caption {The intra-vehicular communication system uses CAN for collision detection system, MOST for audio-video bridging and time-sensitive ethernet networking, FlexRay for the brake-by-wire system, and LIN for multifunctional keyless system~\cite{Renesas:2020}.}
			\label{fig:veh_net}
	\end{center}
	\vspace{-0.6cm}
\end{figure}

The typical automotive data communication networks include twisted wire paired controller area network (CAN), single-wire local interconnect network (LIN), FlexRay, and optical communication-based media oriented systems transport (MOST). Figure~\ref{fig:veh_net} shows the vehicular network data communication systems and some of their example applications. Among all the vehicular data communication networks, CAN is the most popular and widely used in modern automobile systems~\cite{Bosch:1991}.

However, the national highway traffic safety administration (NHTSA) and other researchers identified several security breaches on the vehicular CAN bus~\cite{Nhtsa:2020,Tanksale:2019, Song:2020}. According to the NHTSA, cybersecurity plays a more important role in the vehicular systems than other computational and information systems~\cite{Nhtsa:2020}. In addition to the NHTSA and industry, academic researchers have proposed various intrusion detection systems (IDS) for the CAN bus communications. However, the existing research fails to provide a generalized model that can effectively tackle all kinds of CAN-monitoring-based attacks. We hypothesize that a {\textbf g}raph-based  {\textbf G}aussian 
{\textbf n}aive {\textbf B}ayes (GGNB) IDS for CAN bus to tackle this issue. In particular, the key contributions of this paper are:
\begin{itemize}
    \item This is the first introduction of graph-based model in naive Bayes algorithm for CAN intrusion detection.
    \item A generalized algorithm is proposed that can detect any CAN-monitoring attacks, including mixed attacks without modifying the protocol.
    \item It is the first algorithm that integrates common graph properties with PageRank (PR)-related features into a GNB algorithm.
    \item Overall, the proposed methodology exhibits 98.57\% detection accuracy, which is better or comparable with the state-of-the-art.
    \item The proposed methodology yields many fold reduction in runtime compared to the existing classification methods.
    \item Field-programmable gate array (FPGA) implementation of the proposed GGNB requires $5.7 \times$, $5.9 \times$, $5.1 \times$, and $3.6 \times$ fewer slices, LUTs, flip-flops, and DSP units, respectively, than conventional NN architecture.
\end{itemize}

We organized the rest of the paper as follows: Section~\ref{sec:overview} gives a brief overview
of existing CAN bus attacks and IDSs. 
Section~\ref{sec:proposed_design} proposes our GGNB algorithm. Section~\ref{sec:experiments} compares our proposed IDS with existing schemes. 
Finally, Section~\ref{sec:conclusion} concludes the paper.

\section{Background}
\label{sec:overview}

\subsection{Attacks on CAN Bus}
\label{subsec:attack_CAN}
The ever increasing demand for autonomous vehicles and competition to produce fully autonomous cars by the original equipment manufacturers (OEMs) has grown the CAN bus's security concern. Besides, researchers demonstrated several security vulnerabilities in the CAN bus~\cite{Koscher:2010, Jeong:2021}. This work studies viirtual attacks on the safety-critical systems using a compromised ECU through OBD-II port. The attacker is assumed to have full control of a wide range of functions, for example, disabling the brakes, stopping the engine, and controlling other vehicle functions using reverse engineering code.
Valasek and Miller demonstrated that it was possible to have real-world attacks on multiple vehicles using the CAN bus~\cite{Miller:2015}.  The researchers successfully attacked the brakes of a Jeep Cherokee while it was driving on a highway.
Another interesting approach remotely attacked CAN and showed that it is possible to attack the vehicle's Bluetooth and infotainment systems~\cite{Checkoway:2011}.

\subsection{Existing Intrusion Detection Systems for CAN}
\label{subsec:IDS}

The importance of securing CAN bus led researchers to propose different IDSs~\cite{Moore:2017}.
Moore et al. used the regularity of CAN message frequency to detect the anomaly~\cite{Moore:2017}. A similar detection method proposed by Gmiden et al. relied on the time intervals of CAN messages. Based on the regularity in the signal frequencies, they hypothesized that an accurate detection of a regular-frequency signal injection attack is possible by monitoring the inter-signal wait times of CAN bus traffic.
A highly accurate SVM-based IDS was proposed by Tanksale et al. \cite{Tanksale:2019}. However, it is only applicable to denial of service (DoS) attacks due to the characteristics of the IDS model. 

Another interesting IDS uses a deep convolutional neural network (DCNN) to secure the CAN~\cite{Song:2020}. This method used a frame builder that converts the CAN bus data into a grid-like structure fed into the DCNN. However, this method is not applicable to replay attacks due to the unique characteristics of that attack.
Zhou et al. integrate a deep neural network (DNN) with a triple loss network~\cite{Zhou:2019}. In this method, the researchers extract data features as a set of vectors and then compute the similarity between two CAN data sequences extracted in real-time. Then the integrated approach uses an additional calibrated data sequence to identify the malicious data.

Verendel et al. proposed a honeypot security mechanism at the wireless gateway, acting as a decoy in simulating the in-vehicle networks~\cite{Verendel:2008}. This methodology collected and analyzed the attacked information to update the later version of the security system. The major challenge in deploying honeypot is to make it as realistic as possible and concealing it from intruders.

Tariq et al. proposed a rule-based recurrent neural network (RNN) IDS for CAN~\cite{Tariq:2020}. However, this approach can handle only DoS, fuzzy, and replay attacks. 
On the other hand, Minawi et al. proposed a machine learning-based IDS considering only DoS, Fuzzy, RPM Impersonation, and Gear Impersonation attacks and exhibits good model accuracy~\cite{Minawi:2020}. Similarly, Delwar et al. used only DoS, fuzzy, and spoofing attacks to build their private dataset~\cite{Delwar:2020}. However, unlike these existing methods, we consider mixed attack scenarios. As a result, we removed these methods~\cite{Tariq:2020, Minawi:2020, Delwar:2020} from our analysis and comparisons.


\subsection{Existing Attack Prevention Methodologies}
\label{subsec:prevention}
To secure CAN bus, researchers proposed several techniques~\cite{Lemke:2006,Wang:2018,Islam:2020}. 
Wolf et al.~\cite{Lemke:2006} proposed a firewall signature-based architecture for securing vehicular communication gateways. The firewall only allows authorized controllers to exchange valid messages. However, this method can not fully shield the vehicle network because most modern vehicles have interfaces that enable access to the entire car system.
Islam et al.~\cite{Islam:2020} proposed the first dynamic arbitration ID refreshing algorithm for preventing CAN traffic monitoring based attacks. They validated their methodology using real CAN and implementing virtual CAN to demonstrate their algorithm's efficiency.

\subsection{Existing Graph-Based Anomaly Detection}
\label{subsec:graph_anomaly}
In recent years, we observed a significant increase in graph-based anomaly detection due to the ease of representation of complex data using graphs. Researchers used background knowledge of the evaluation metrics and biased the substructure discovery process towards discovering anomalous substructures~\cite{Velampalli:2017}. Background knowledge is added in the form of rule coverage, which conveys the percentage of the final graph covered by the substructure instances. The authors hypothesized that it is possible to discover anomalous substructures by introducing negative weights to the rule coverage. This method can efficiently detect 100\% of the attacks with no false positives rate using the KDD99 data set. Another relevant approach uses graph-based anomaly detection by analyzing the data for the suspicious employee activities at Kasios~\cite{Velampalli:2019}. This approach provides a rich contextual and more in-depth understanding of data due to
its ability to discover patterns in databases which are not a trivial task to identify using traditional query or statistical tools. This method achieves a true positive rate and false positive rate of 100\% and 65\%, respectively.

\subsection{Types of Attacks on CAN}
\label{subsec:types_attacks}
The electronic fusion of automobiles brings many benefits, including driver comforts, performance, and vehicle efficiency. However, these benefits comes at a higher risk of cyber attacks and human safety due to automobiles' dependencies on those devices. Researchers and OEMs already identified security threats in CAN communication and agreed that it was not built with security in mind~\cite{Ueda:2015,Studnia:2013,Carsten:2015,Boudguiga:2016,Staggs:2013, Hoppe:2011}. In this research, we consider both fabrication and masquerade CAN attacks. When an intruder transmits messages using a compromised ECU with a forged ID is referred to as a fabrication attack. DoS, fuzzy, and spoofing attacks are also classified under fabrication attacks~\cite{Ueda:2015, Staggs:2013, Boudguiga:2016, Hoppe:2011}. 
The diagnostic attack consists of injecting messages of CAN ID which values are in a particular range using a diagnostic tool~\cite{Woo:2014, Guillaume:2019}. 
Fuzzy attacks are mainly composed of fuzzing CAN ID and fuzzing payload attacks~\cite{Lee:2017, Guillaume:2019}. The fuzzing CAN ID  consists of injecting messages with a specific CAN ID. These CAN ID values are not a part of the legitimate values. Fuzzing payload attack involves modifying the payload of messages that are not used in legitimate traffic. Another weak attack, where hackers sniff in-vehicle networks and suspend message transmission of a certain ECU, is called suspension attack~\cite{Haojie:2018, Guillaume:2019}.
On the other hand, 
a masquerade attack requires two compromised ECUs to mount an attack. The masquerade attacks are considered strong attacks. A replay attack is known as a masquerade attack~\cite{Studnia:2013}.

\section{Proposed Anomaly Detection Methodology} 
\label{sec:proposed_design}

In this paper, we propose a graph-based anomaly detection system to secure the CAN bus communication system. The proposed methodology is based on the PR analysis for detecting anomalies. PR is a well-known algorithm used by Google Search Engine to rank their websites~\cite{Page:1999}. The PR algorithm can effectively suggest users for searching different topics in Google as their requirements. Before presenting the proposed anomaly detection algorithm, we will discuss the PR algorithm.

\subsection{PageRank Algorithm} 
\label{subsec:pagerank}
\begin{figure}[t]
	\begin{center}
		\vspace{-0.25cm} 
		\includegraphics[width = 0.5\textwidth]{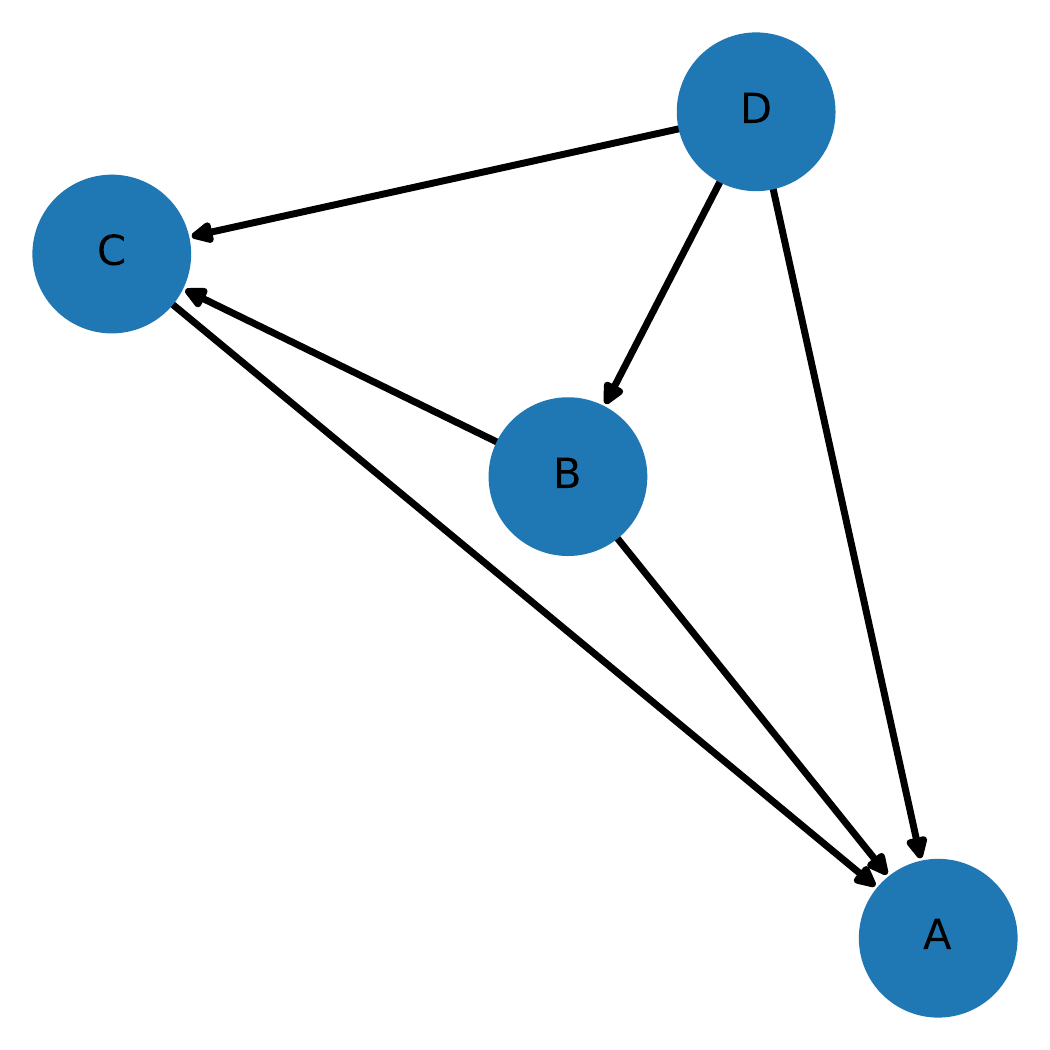}
		\vspace{-0.25cm} 
		\caption {Example four-vertex directed graph for PR computation.}
			\label{fig:example_graph}
	\end{center}
	\vspace{-0.1cm}
\end{figure}

We will discuss the PR algorithm using a simple graph with four vertices A, B, C, and D, as shown in Figure~\ref{fig:example_graph}.
PR is initialized to the same value for all vertices, which assigns a probability distribution between 0 and 1. So, the initial probability distribution of each C is $\frac{1}{n}=0.25$, where n is the total number of vertices.

Every vertex will contribute PR to other vertices with the outgoing edges. As a result, vertex D will contribute to A, B, and C vertices. Since D has three outlines, it will divide its PR equally into three parts. Here, we compute the PR of A as,
\begin{equation}
PR(A) = \frac{PR(B)}{Outdegree(B)} + \frac{PR(C)}{Outdegree(C)} + \frac{PR(D)}{Outdegree(D)}
\label{eq:PR}
\end{equation}
where Outdegree(B) is the number of edges exiting from B. We iteratively compute PR of each vertex and assign the PR value. The iteration will stop when there will be no update of PR of vertices.
We computed the PR values of vertices for both attacked and attack-free graphs. To identify the significance of PR values, we performed an analysis of many graphs. According to our observation, PR plays a vital role in detecting anomalies. Empirically, the distribution of PR values among attack-free vertices is significantly different from the attacked vertices. In an attack-free graph, the distribution of PR among different vertices is similar, whereas, in an attacked graph, the PR values of attacked vertices are significantly different from other vertices in the graph. As a result, we used PR-related features for our anomaly detection algorithm.

\begin{figure}
    \centering
    \vspace{-0.5cm}
    \subfigure[]{\includegraphics[width=0.4\textwidth]{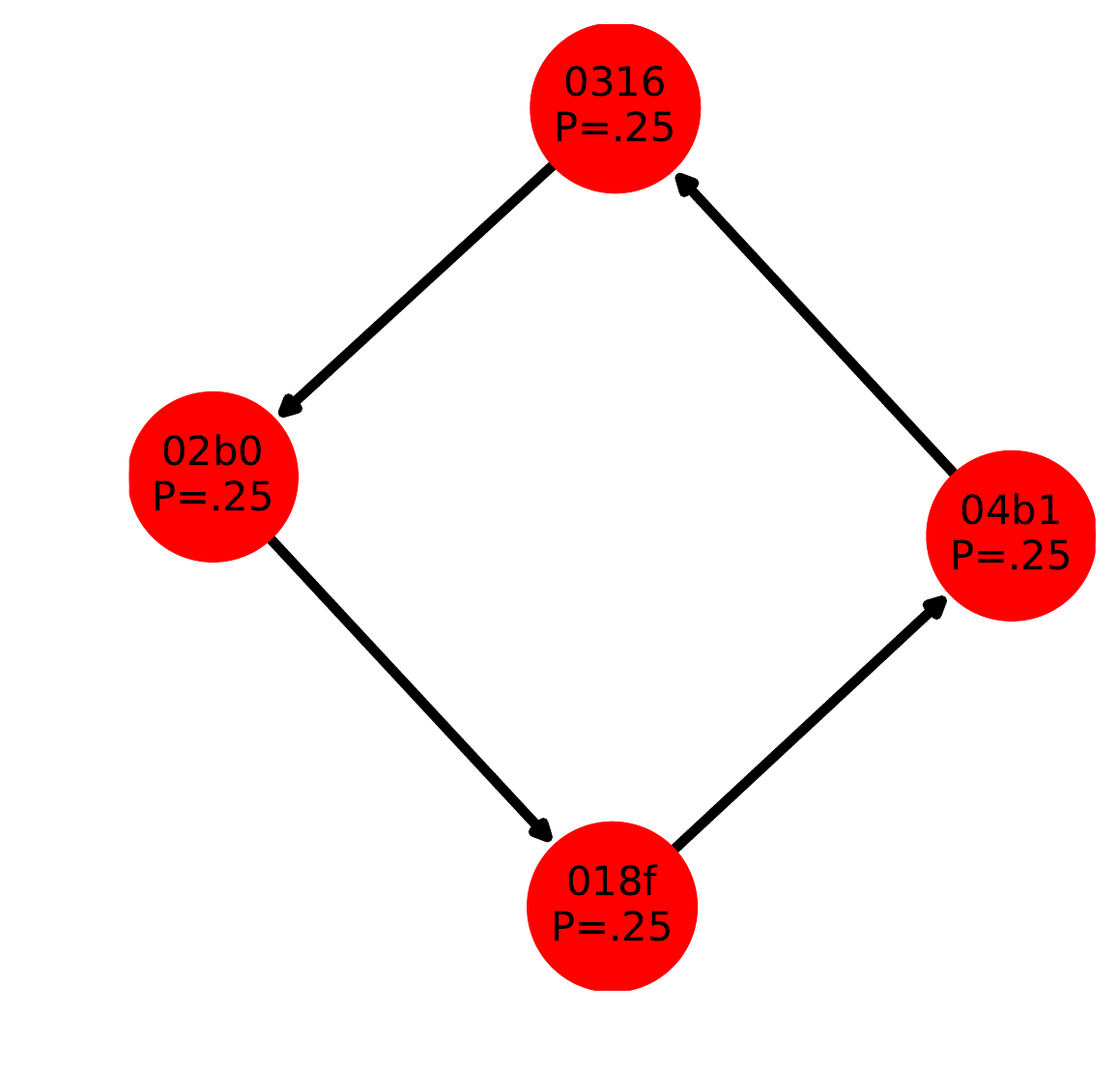}} 
    \subfigure[]{\includegraphics[width=0.4\textwidth]{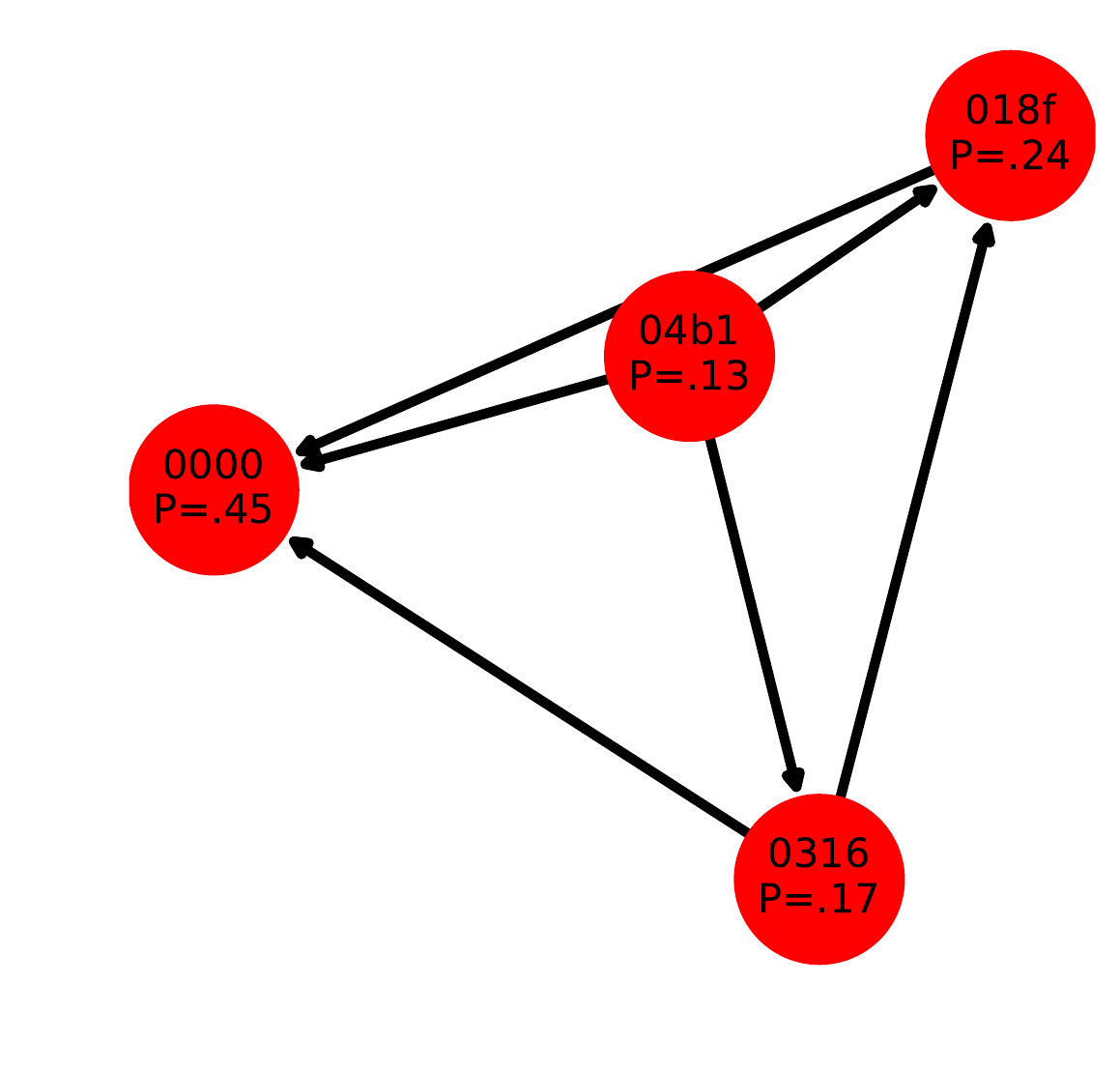}} 
    \caption{(a) A simple graph where each vertex has an equal 0.25 PR value, and (b) when a DoS attack is introduced, the PR value of 0000 ID is significantly higher than the other vertices.}
    \label{fig:dos_example}
    \vspace{-0.15cm}
\end{figure}

When an intruder mounts a DoS attack on a network, it floods the communication network with a certain ID with high priority. We depict a similar situation in Figure~\ref{fig:dos_example}, where Figure~\ref{fig:dos_example}(a) shows a simple graph with each vertex with equal 0.25 PR value. On the other hand, Figure~\ref{fig:dos_example}(b) shows an example when vertex ID with 0000 has a PR value of 0.45, which is significantly higher than the other vertices of that graph.

\begin{algorithm}
	\begin{small}
		\caption{PageRank-related features extraction algorithm}
		\label{alg:PR}
		\begin{algorithmic}[1]
			\State {\bf Input:} G, collection of graphs; \\ 					
			{\bf Output:} The list of minimum, median, and maximum PR of each graph in G;
			\State
			\State $graphList = [ ]$ \Comment{initializing the list of the PR of every vertex from each graph}\label{alg1:line4}
			\State $graphListWithMedianPR = [ ]$ \Comment{initializing the list of the median PR from each graph} \label{alg1:line7} 
			
			\State $graphListWithMaximumPR = [ ]$ \Comment{initializing the list of the maximum PR from each graph} \label{alg1:line9i}
			\State $graphListWithMinimumPR = [ ]$ \Comment{initializing the list of the minimum PR from each graph} \label{alg1:line9}
\ForAll{$g_i \in G:$} \label{alg1:line10}
\State{$graphList[i] = PR(g_i)$} \Comment{$PR(g_i)$ computes the PR of every vertex of $g_i$} 
\label{alg1:line11}
\State{$graphListWithMedianPR[i] = median(graphList[g_i])$} \label{alg1:line12}
\State{$graphListWithMaximumPR[i] = maximum(graphList[g_i])$}  \label{alg1:line13}
\State{$graphListWithMinimumPR[i] = minimum(graphList[g_i])$} \label{alg1:line14}
\EndFor \label{alg1:line15}
			
		\end{algorithmic}
	\end{small}
\end{algorithm}

Algorithm~\ref{alg:PR} shows the PR and other PR-related feature computation steps. We initialize all the necessary variables from Line~\ref{alg1:line4} to  Line~\ref{alg1:line9}. Then we iteratively compute the minimum, median, and maximum 
PR values of every graph $g_i \in G$ in Line~\ref{alg1:line10} to Line~\ref{alg1:line15}. 	

\subsection{Proposed Graph-Based Gaussian Naive Bayes (GGNB) IDS} 
\label{subsec:gnb}
The proposed graph-based IDS uses existing Gaussian naive Bayes (GNB) classifier~\cite{John:2013}. The GNB  follows Gaussian normal distribution and supports continuous data. 
In general, naive Bayes uses the posterior probability of Bayesian rule
to infer a classification label from input feature distribution.
The conditional posterior probability can be defined as the probability of an event (A) occurring given the probability of another event that has already occurred (B) and can be expressed as
\begin{equation}
P(A \mid B) = \frac{P (A \cap B)}{ P(B)} = \frac{P (A)\cdot P(B \mid A)}{ P(B)}
\label{eq:cond_prob}
\end{equation}
where P(A) and P(B) are distributions of events A and B, respectively. GNB classifiers assume that features are independent of each other and require a small training data to estimate the parameters needed for classification. Since we assume that the likelihood of the features to be Gaussian. So, the likelihood probability is given by:

\begin{equation}
P(x_{i}\mid y = y') = \frac{1}{\sqrt{2\pi \sigma_x^{2}}} \exp \left(-\frac{(x_{i} -\mu_{x})^2}{2\sigma_x^{2}} \right)
\end{equation}

where $\mu$ is the mean, and $\sigma$ is the standard deviation. Assume that we have an unknown binary outcome $Y$ and $n$ continuous features of $X$. As we have a binary label, $Y$  is Bernoulli, which can be defined by setting a positive probability.  The likelihood of $f_{X|Y}(x|y)$ is the distribution of the observation for the known outcome. Our updated knowledge or posterior about the unknown after observation is $p_{Y|X}(y|x)$. We can express the Bayes rule as

\begin{equation}
p_{Y|X}(y|x)=\frac{p_Y(y)f_{X|Y}(x|y)}{\sum_{y'}p_Y(y')f_{X|Y}(x|y')}
\label{eq:bayes_rule}
\end{equation}
The Equation~\ref{eq:bayes_rule} is true for the $n$ continuous variables $X = X_0,X_1,\ldots,X_{n-1}$ and can be expressed as,

\begin{equation}
\begin{multlined}
p_{Y|X_0,X_1,\ldots,X_{n-1}}(y|x_0,x_1,\ldots,x_{n-1})= \\
\frac{p_Y(y)\prod_{i=0}^{n-1}f_{X_i|Y}(x_i|y)}{\sum_{y'=0}^1p_Y(y')\prod_{i=0}^{n-1}f_{X_i|Y}(x_i|y')}
\label{eq:bayes_series_rule}
\end{multlined}
\end{equation}

Before discussing the proposed algorithm, we give a simple example of an attack detection mechanism. Assume that in a certain number of graphs, we have the probability of attacked and attack free graphs are $0.5$. Then we compute the normal distribution of each feature of those graphs. To identify a graph with $V_i$ vertices and $E_i$ edges weather it is attacked or attack free, we utilized the likelihood (L) condition as 
\begin{equation}
sAtt = P(att)*L( Nodes = V_i \mid att)* L( Edges = E_i \mid att)
\label{eq:likelihood1}
\end{equation}
\begin{equation}
\begin{multlined}
sAttFr = P(attFr)*L( Nodes = V_i  \mid attFr)* \\
L( Edges = E_i\mid attFr)
\label{eq:likelihood2}
\end{multlined}
\end{equation}

where $P(att)$ and $P(attFr)$ are probability of attack and attack free data, respectively.  And If $sAtt$ and $sAttFr$ values are underflow, a $\log$ function before them can help us to restore a comparable value. The algorithm compares the $sAtt$ and $sAttFr$ values to decide if the graph is attacked or attack free. If $sAtt$ is greater than $sAttFr$ then the graph is attacked. Table~\ref{tab:features} shows the list of features we used in our proposed anomaly detection algorithm.
\begin{table}[t!] \large 
\vspace{-0.15cm}
\renewcommand{\arraystretch}{1.5}
\captionsetup{aboveskip=-0.00cm,belowskip=-0.25cm}
\caption{
		The proposed methodology uses graphs' vertices, edges, degrees, and PageRank-related features for classification.
		\label{tab:features}}
\centering
\scalebox{0.65}{
\begin{tabular}{|c|c|c|c|c|c|c|c|c|c|c|c|c|c|c|r|}
 \hline
 Features & Names  
 \tabularnewline
\cline{1-2}
 Number of nodes in a graph & Nodes or Vertex
 \tabularnewline
\cline{1-2}
  Number of edges in a graph & Edges 
  \tabularnewline
\cline{1-2}
 Maximum number of in-degree in a graph & Maximum Indegree   
  \tabularnewline
\cline{1-2}
 Maximum number of out-degree in a graph & Maximum Outdegree 
  \tabularnewline
\cline{1-2}
 Minimum number of in-degree in a graph & Minimum Indegree 
  \tabularnewline
\cline{1-2}
  Minimum number of out-degree in a graph & Minimum Outdegree  
  \tabularnewline
\cline{1-2}
   Median of PageRanks among nodes in a graph & Median PageRank  
  \tabularnewline
\cline{1-2}
 Maximum of PageRanks among nodes in a graph & Maximum PageRank 
 \tabularnewline
\cline{1-2}
  Minimum of PageRanks among nodes in a graph & Minimum PageRank

\tabularnewline
\cline{1-2}

\end{tabular}}
\vspace{-0.0cm}
\end{table}

\begin{algorithm}
	\begin{small}
		\caption{Anomaly Detection Algorithm}
		\label{alg:anomaly_detect}
		\begin{algorithmic}[1]
			\State {\bf Input:} Raw CAN data, WindowSize; \\ 					
			{\bf Output:} Prediction; \Comment{Attacked if Prediction is correct else benign}
			\State
			\State $G = CreateGraph(Raw CAN data, WindowSize)$ \label{alg2:line4}
			\State $graphListWithMaximumIndegree = [ ]$ \Comment{initializing the list of maximum in-degree from each graph}\label{alg2:line5}
			\State $graphListWithMaximumOutdegree = [ ]$ \Comment{initializing the list of maximum out-degree from each graph}\label{alg2:line6}
			\State $graphListWithMinimumIndegree = [ ]$ \Comment{initializing the list of minimum in-degree from each graph}\label{alg2:line7}
\State $graphListWithMinimumOutdegree = [ ]$ \Comment{initializing the list of minimum out-degree from each graph}\label{alg2:line8}
\State $graphListWithNodesNumber = [ ]$ \Comment{initializing the list of number of nodes from each graph}\label{alg2:line9}
\State $graphListWithEdgesNumber = [ ]$ \Comment{initializing the list of number of edges from each graph}\label{alg2:line10}
			
			\ForAll{$g_i \in G:$} \label{alg2:line11}
\State{$graphListWithNodesNumber[i] = numberOfNodes(g_i)$} \label{alg2:line12}
\State{$graphListWithEdgesNumber[i] = numberOfEdges(g_i)$} \label{alg2:line13}
\State{$graphListWithMaximumIndegree[i] = maximumIndegree(g_i)$}  \label{alg2:line14}
\State{$graphListWithMaximumOutdegree[i] = maximumOutdegree(g_i)$} \label{alg2:line15}
\State{$graphListWithMinimumIndegree[i] = minimumIndegree(g_i)$}  \label{alg2:line16}
\State{$graphListWithMinimumOutdegree[i] = minimumOutdegree(g_i)$} \label{alg2:line17}
\EndFor \label{alg2:line18}
			\State $\{PageRankFeaturesList\}  = GraphListPageRank(G)$ \Comment{extract PR-related feature using Algorithm~\ref{alg:PR}} \label{alg2:line19}
			\State $\{data, labels\} = processData()$ \label{alg2:line20}
			\State $train, test, trainLabels, testLabels = trainTestSplit(data,labels)$ \Comment{splitting the data and labels for training and testing} \label{alg2:line21} 
			\State $model = gaussianNb(train, trainLabels)$ \Comment{training data} \label{alg2:line22} 
			\State $predictions = model.predict(test)$ \Comment{making the predictions} \label{alg2:line23} 
		\end{algorithmic}
	\end{small}
\end{algorithm}

The proposed anomaly detection methodology is shown in Algorithm~\ref{alg:anomaly_detect}. It accepts raw CAN data and window size as inputs and predicts whether the input is attacked or attack free. For the analysis, we used $\sim 23ms$ as window size considering 1Mbit/s CAN speed. The proposed algorithm will first create a list of graphs $g_{1},g_{2}, \ldots, g_{n}$ named $G$ based on the user-provided window size in Line~\ref{alg2:line4}. The CAN message IDs will act as vertices, and we construct an edge between two vertices depending on their sequence. We declared different list variables from Line~\ref{alg2:line5} to Line~\ref{alg2:line10}. In Line~\ref{alg2:line11}, we declared a for loop. In each iteration of the for loop, we collect nodes, edges, maximum in-degree, maximum out-degree, minimum in-degree, minimum out-degree from each graph $g$ from Line~\ref{alg2:line10} to Line~\ref{alg2:line18}. Then we compute the PR-related features using the $GraphListPageRank(G)$ function which calls the Algorithm~\ref{alg:PR} in Line~\ref{alg2:line19}. The data processing and labeling are performed in Line~\ref{alg2:line19}.
In between Line~\ref{alg2:line21} and Line~\ref{alg2:line22}, we split the data and labels into training and test sets and fit the data into a GNB model. We perform attack prediction in Line~\ref{alg2:line23}.

\section{Experiments}
\label{sec:experiments}
For verifying the proposed methodology, we used real CAN data and performed analysis on an Intel Xeon(R) Silver 3.2GHz 20-core processor with 48GB RAM by using our proposed methods in Python language. In our research, we used raw CAN attacked and attack free data~\cite{Lee:2017}. We refer to this data set as rawCAN. The attacked data includes DoS, fuzzy, spoofing, and replay attacks. Besides, we consider the mixed attacks that have all four kinds of attacks. From the raw CAN bus data, we build $\sim 18.5K$, $\sim 37K$, $\sim 38K$, $\sim 42K$, and $\sim 19K$, and $\sim 80K$ attack free, DoS, fuzzy, spoofing, replay, and mixed attacked graphs. 

\begin{figure*}
\captionsetup{aboveskip=-0.00cm,belowskip=-0.050cm}
    \centering
    \vspace{-0.5cm}
    \subfigure[]{\includegraphics[width=0.25\textwidth]{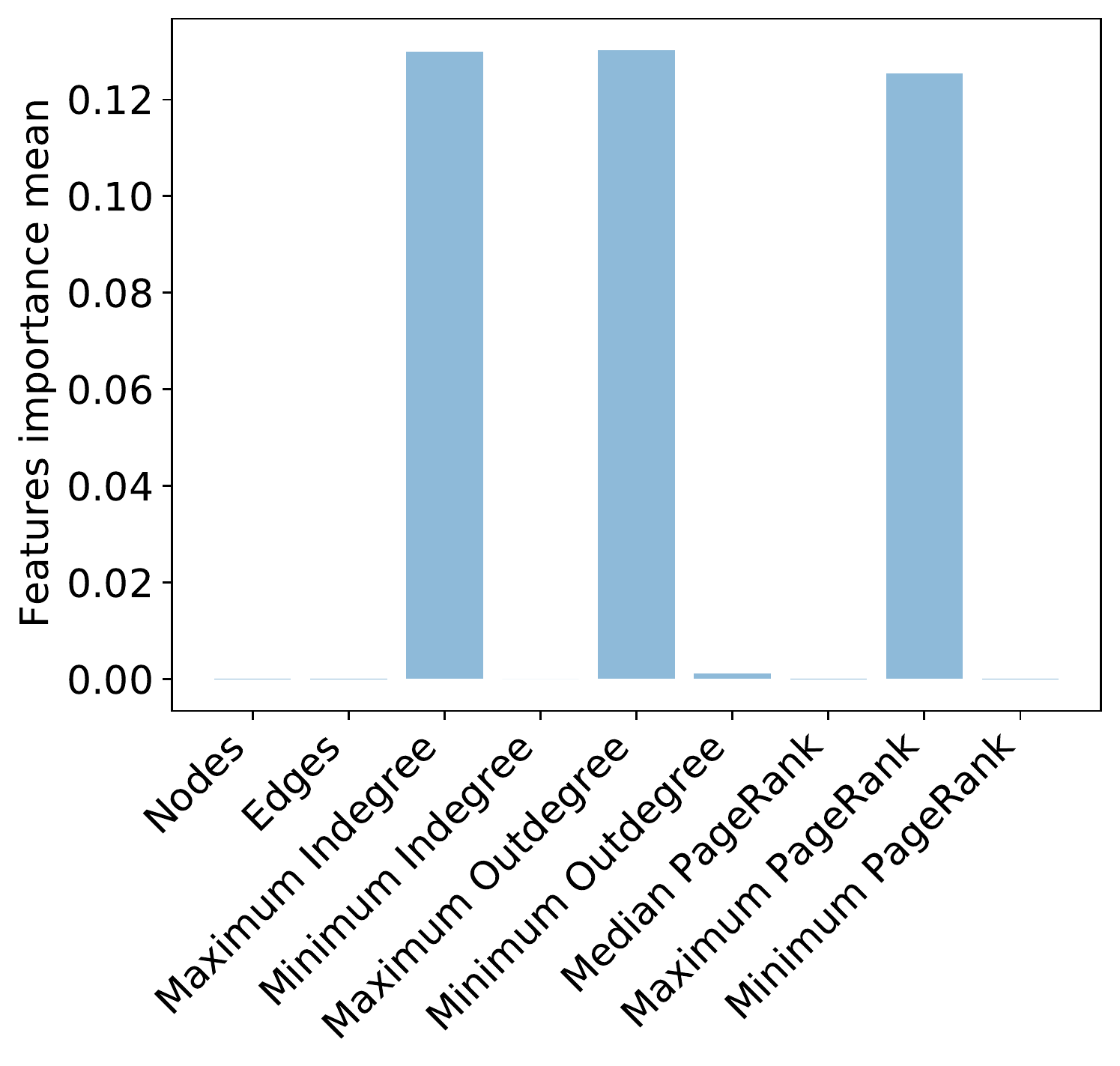}} 
    \subfigure[]{\includegraphics[width=0.25\textwidth]{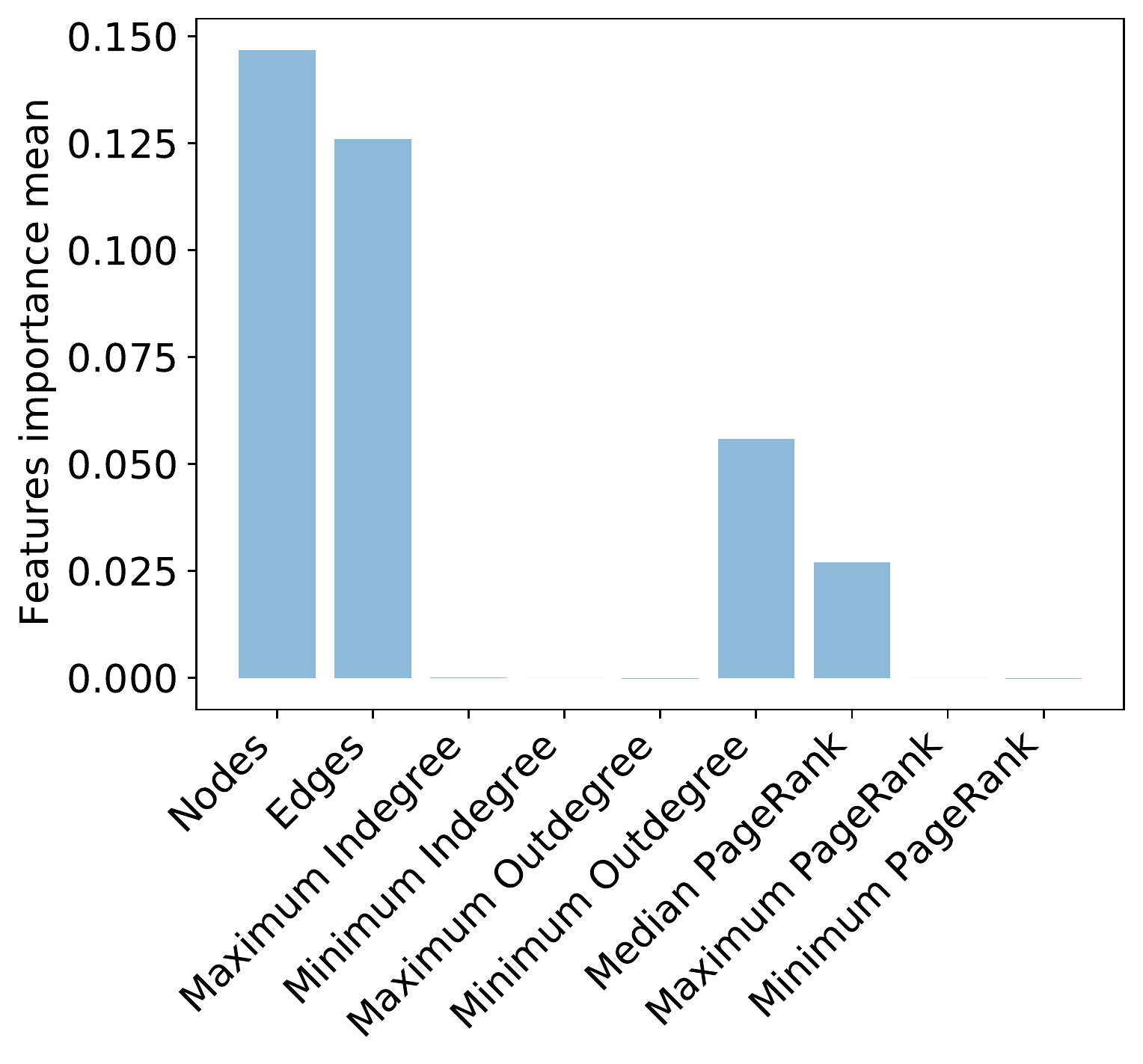}} 
    \subfigure[]{\includegraphics[width=0.25\textwidth]{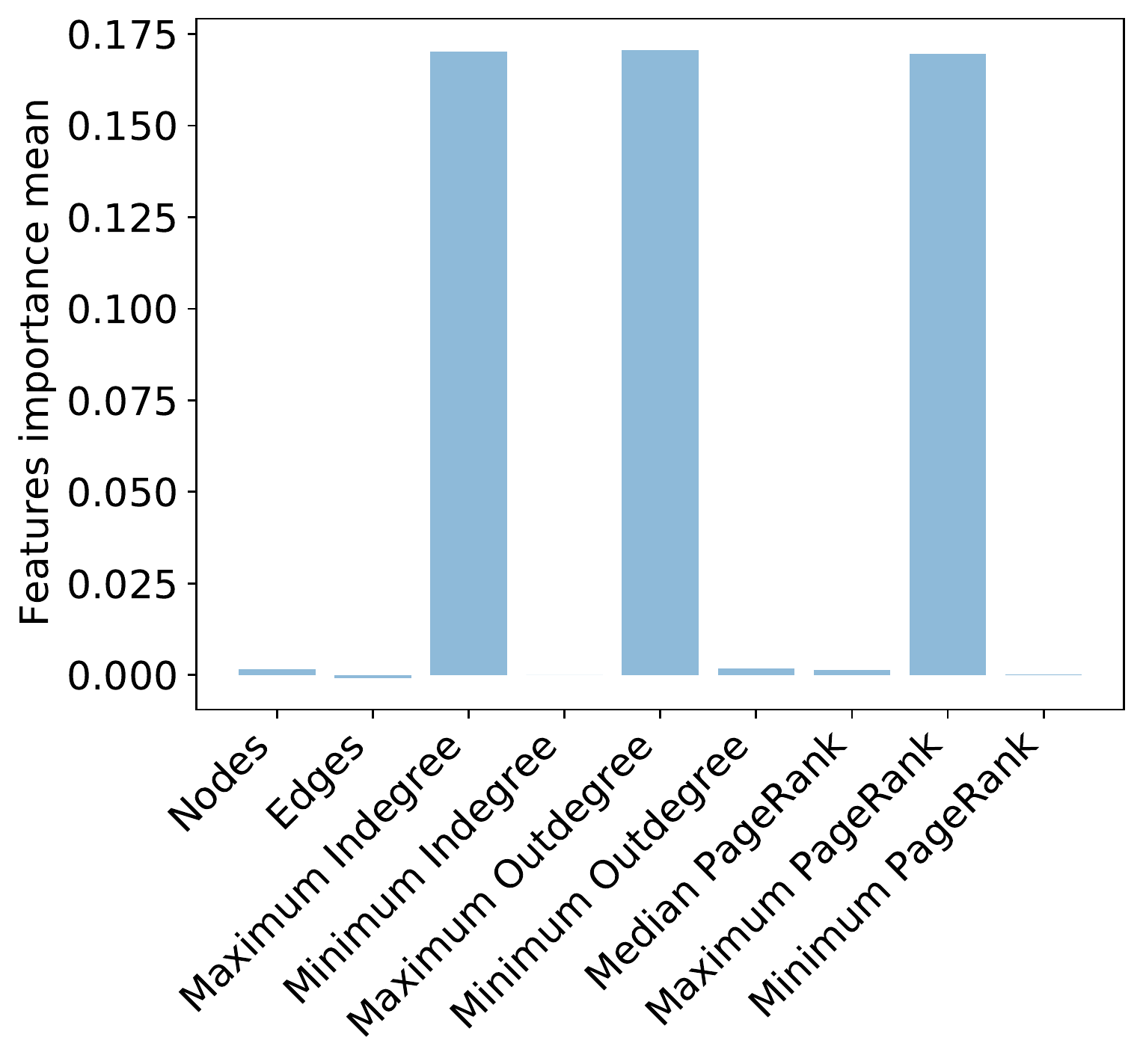}}
    \subfigure[]{\includegraphics[width=0.25\textwidth]{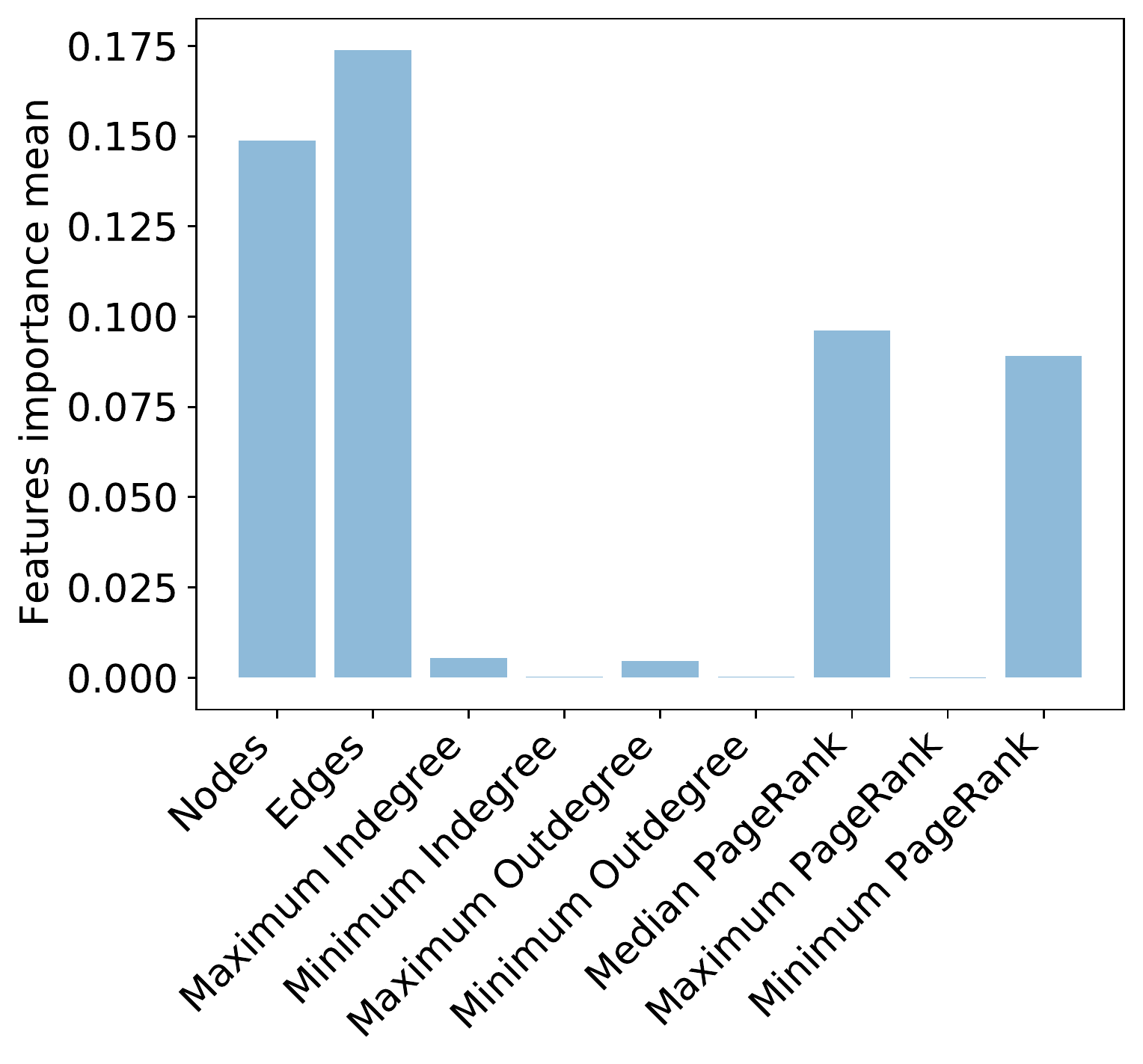}}
    \subfigure[]{\includegraphics[width=0.25\textwidth]{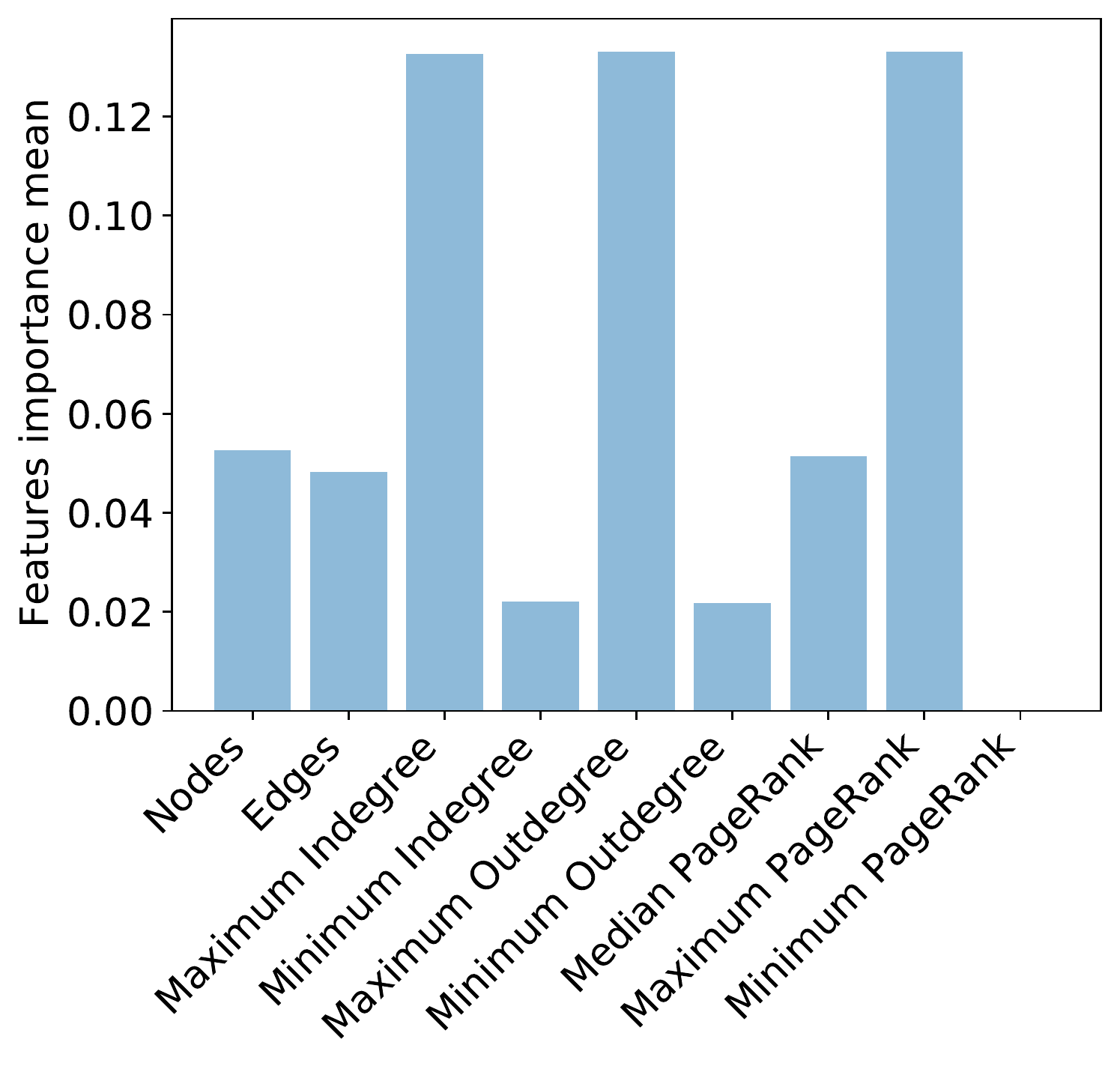}}
    \caption{The Maximum Indegree, Maximum Outdegree, and Maximum PageRank have the highest feature importance for DoS, spoofing, and mixed attacks, respectively and nodes and edges have the highest feature importance for fuzzy and replay attacks.}
    \label{fig:feature_imp_all}
    \vspace{-0.1cm}
\end{figure*}

To apply the proposed GGNB-based anomaly detection algorithm, we used standard graph features listed in Table~\ref{tab:features}. Besides, we identify PR-related features to help detect anomalies. It primarily helps removing the non-informative features from the model \cite{Kuhn:2013}. We computed feature importance for DoS, fuzzy, spoofing, replay, and mixed attacked data, as shown in Figure~\ref{fig:feature_imp_all}.

\begin{figure}
\captionsetup{aboveskip=-0.00cm,belowskip=-0.050cm}
    \centering
    \vspace{-0.05cm}
    \subfigure[]{\includegraphics[width=0.4\textwidth]{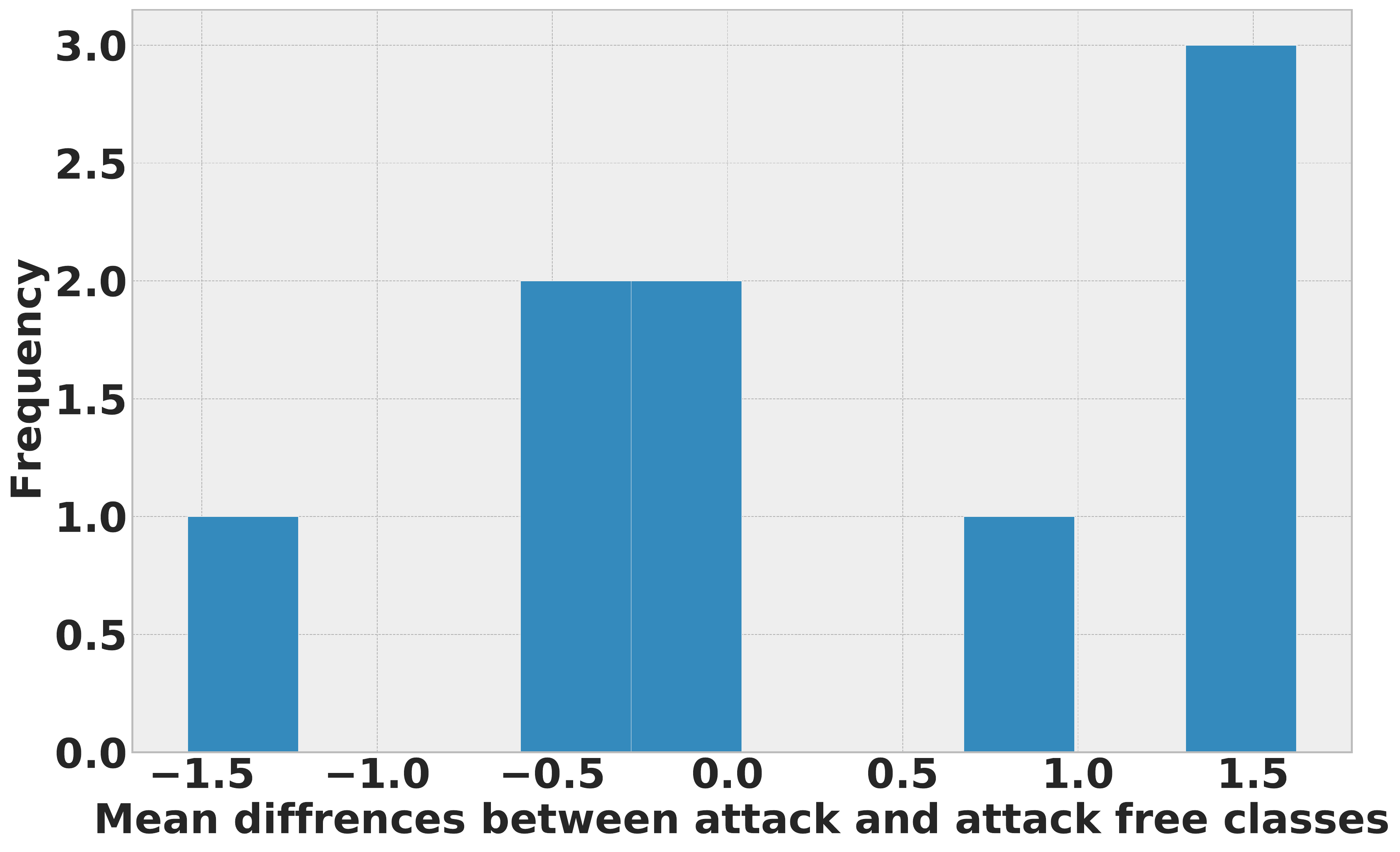}} 
    \subfigure[]{\includegraphics[width=0.4\textwidth]{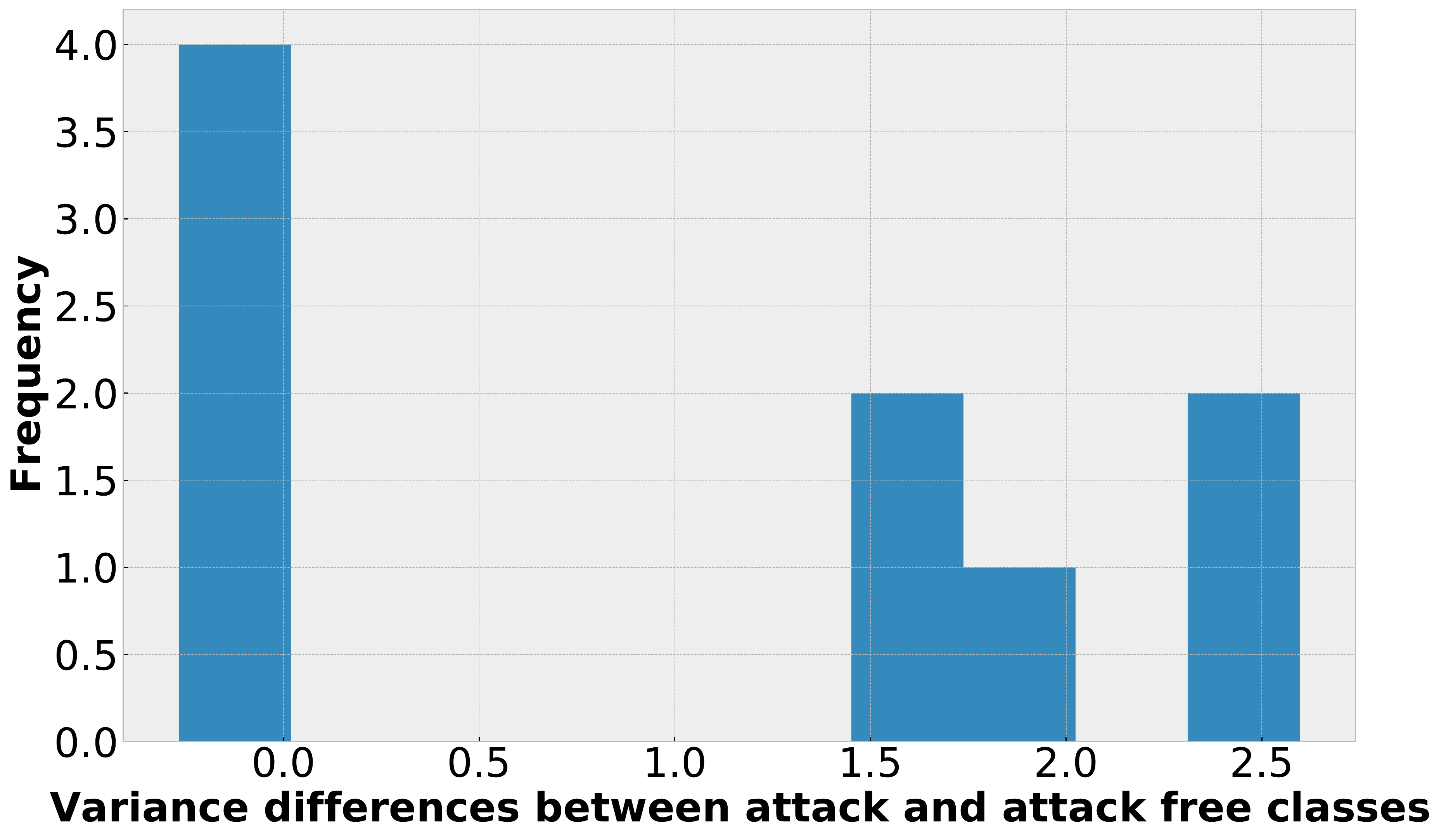}} 
    \caption{(a) The mean differences between the two classes have symmetric spread in both positive and negative direction, and (b) the variance differences of the feature samples are entirely on the positive side.}
    \label{fig:mean_var_diff}
    \vspace{-0.05cm}
\end{figure}

We compute the mean differences of the sample features after quantile transformation, which shows that distinctions are balanced around zero, as shown in Figure~\ref{fig:mean_var_diff}(a).
However, the variance differences of the feature samples are entirely on the positive side. So, Figure~\ref{fig:mean_var_diff}(b) depicts that the positive likelihood distributions are more concentrated around their means than the negative ones. These indicate the discriminating power of the model. We can classify data if the mean difference is far away from the distributions' centers, or the variance difference is greater than the distributions' spread.

\begin{figure}
\captionsetup{aboveskip=-0.00cm,belowskip=-0.050cm}
    \centering
    \vspace{-0.15cm}
    \subfigure[]{\includegraphics[width=0.4\textwidth]{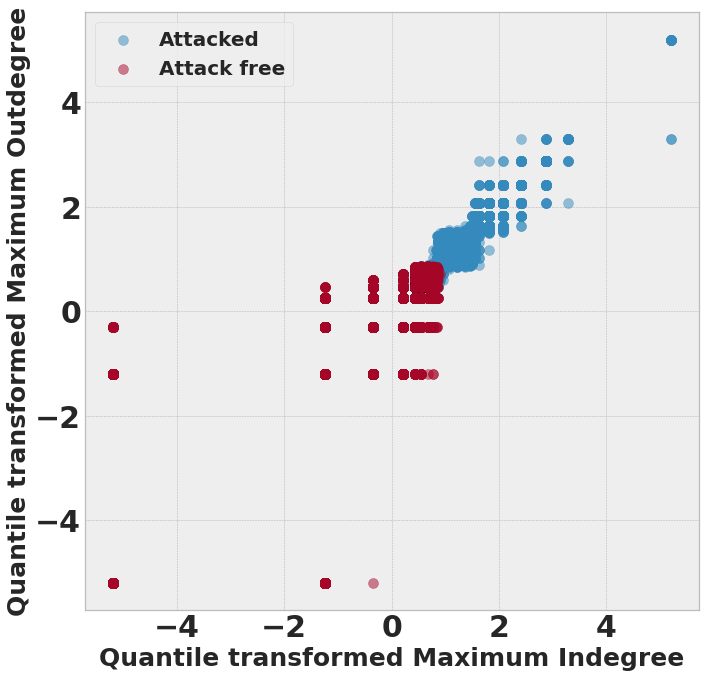}} 
    \subfigure[]{\includegraphics[width=0.4\textwidth]{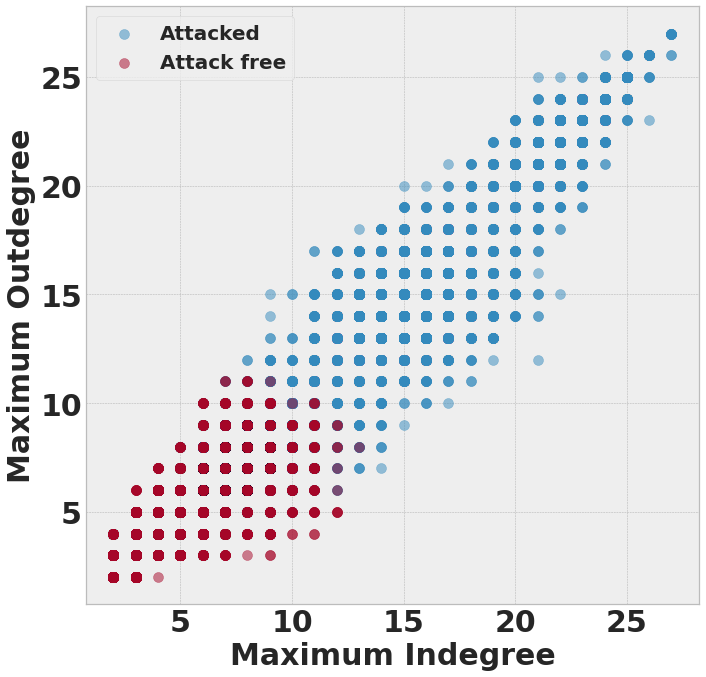}} 
    \caption{Separation of classes using Maximum Indegree and Maximum Outdegree features considering (a) quantile transformed feature, and (b) original features values of mixed attacks.}
    \label{fig:quantile_tranform}
    \vspace{-0.60cm}
\end{figure}

After quantile transformation, we identified two features, Maximum Indegree and Maximum Outdegree, which distinguish the two classes very well. To detect positive or negative class, we observed a closeness between the transformed values of these two features for any graph. Hence, a two-dimensional plot using the mixed attacked data provides a straight line through the origin, as shown in Figure~\ref{fig:quantile_tranform}. The transformed features will maintain a range of values for most of the graphs of each class. So, after plotting altered Maximum Indegree and Maximum Outdegree features, we observe a good separation between negative and positive class, as shown in Figure~\ref{fig:quantile_tranform}(a). Even without quantile transformation, the original features' values can separate the classes efficiently, as demonstrated in Figure~\ref{fig:quantile_tranform}(b). Hence, using these features, the proposed methodology works well for detecting attacked and attack free graphs.

\begin{figure}[t]
	\begin{center}
		\vspace{-0.0cm} 
		\includegraphics[width = 0.5\textwidth]{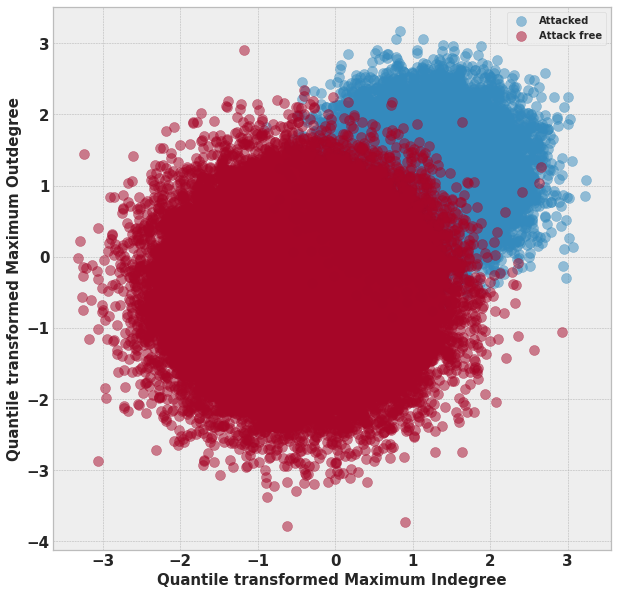}
		\vspace{-0.4cm} 
		\caption {The quantile transformed features values in the range from $\mu$-to-$\pm 3\sigma$ can clearly separate attack and attack-free data.}
			\label{fig:quantile_tranform_mu_sigma}
	\end{center}
	\vspace{-0.2cm}
\end{figure}

The quantile transformation of the features helped us identify the classes. Then, we compute the transformed features' mean and standard deviations for both positive and negative categories.  For the transformed Maximum Indegree and Maximum Outdegree, we select random values in the range from $\mu$-to-$\pm 3\sigma$. When we draw the resulting values, it can clearly distinguish two classes, as shown in Figure~\ref{fig:quantile_tranform_mu_sigma}.

\subsection{Comparison of Prediction Accuracy}
\label{subsec:accuracy}
\begin{figure*}
\captionsetup{aboveskip=-0.00cm,belowskip=-0.250cm}
    \centering
    \vspace{-0.5cm}
    \subfigure[]{\includegraphics[width=0.3\textwidth]{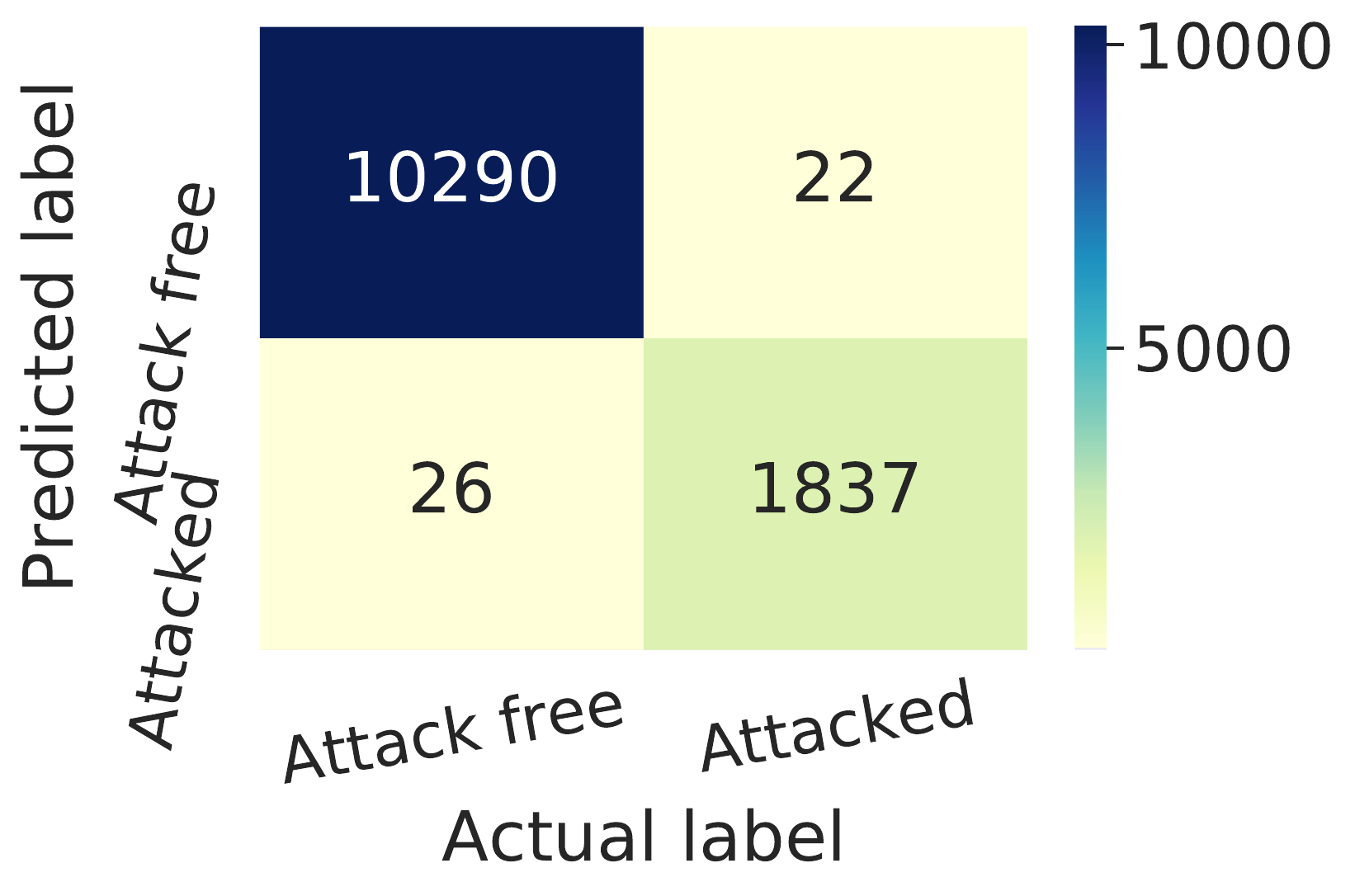}} 
    \subfigure[]{\includegraphics[width=0.3\textwidth]{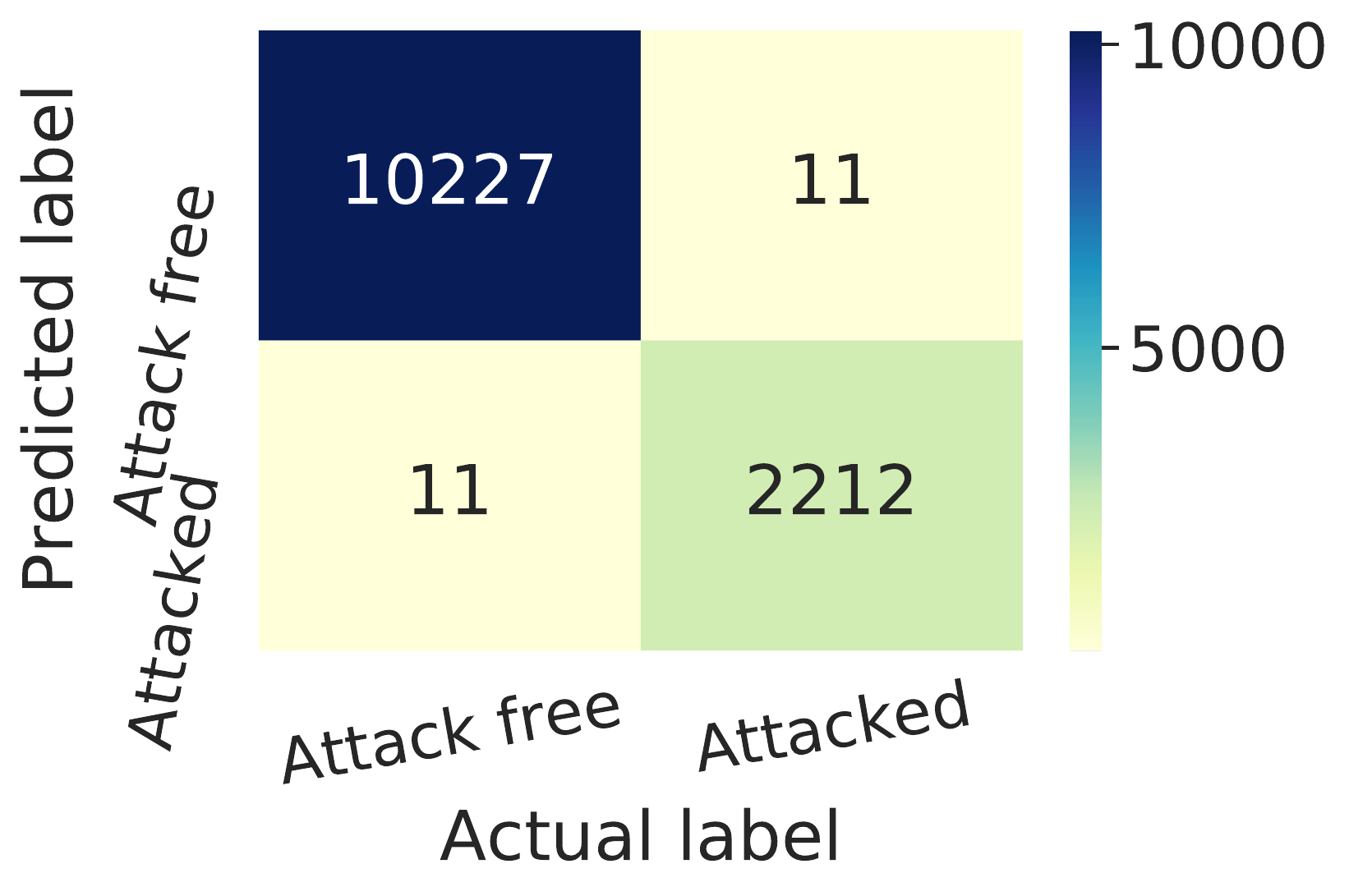}} 
    \subfigure[]{\includegraphics[width=0.3\textwidth]{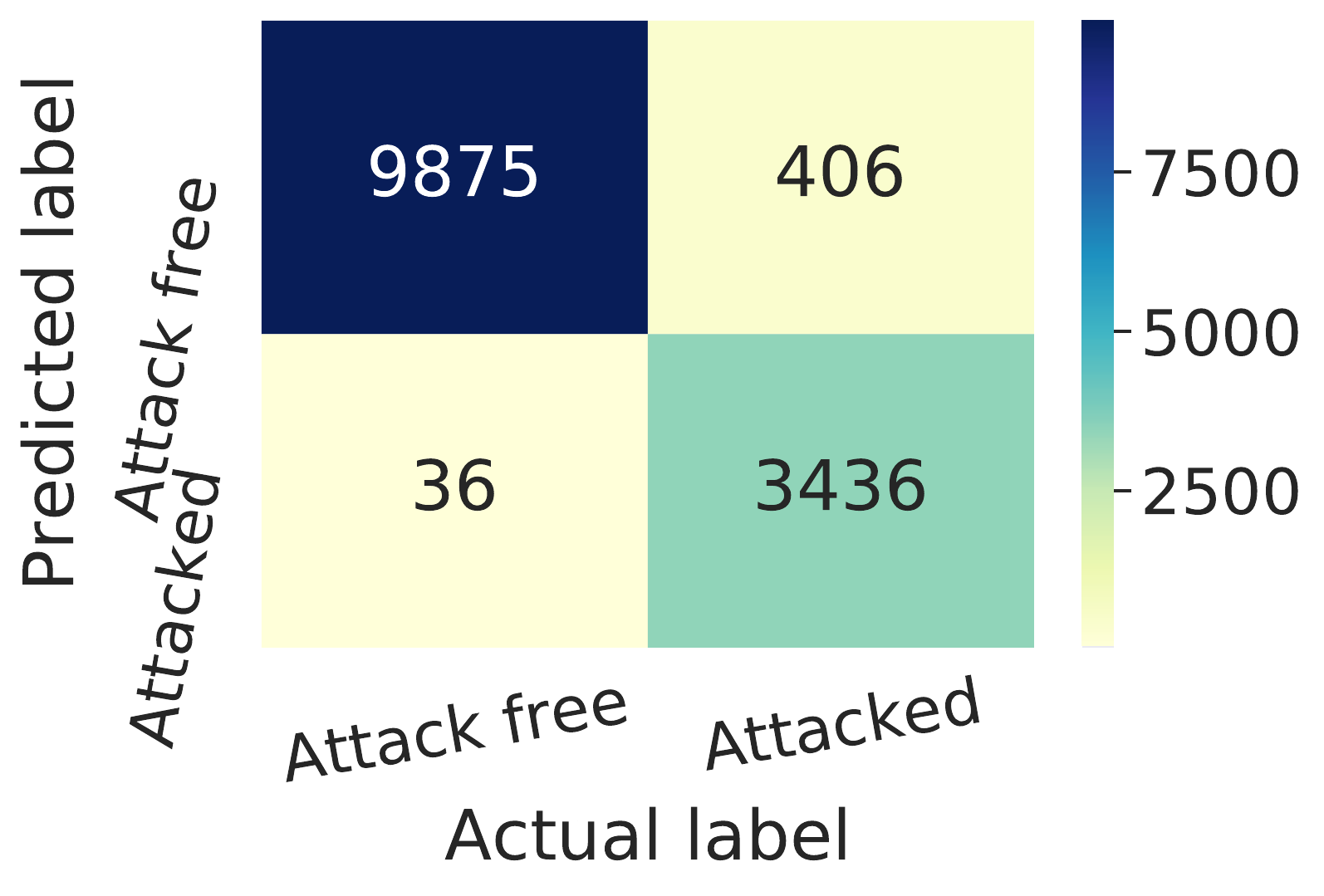}}
    \subfigure[]{\includegraphics[width=0.3\textwidth]{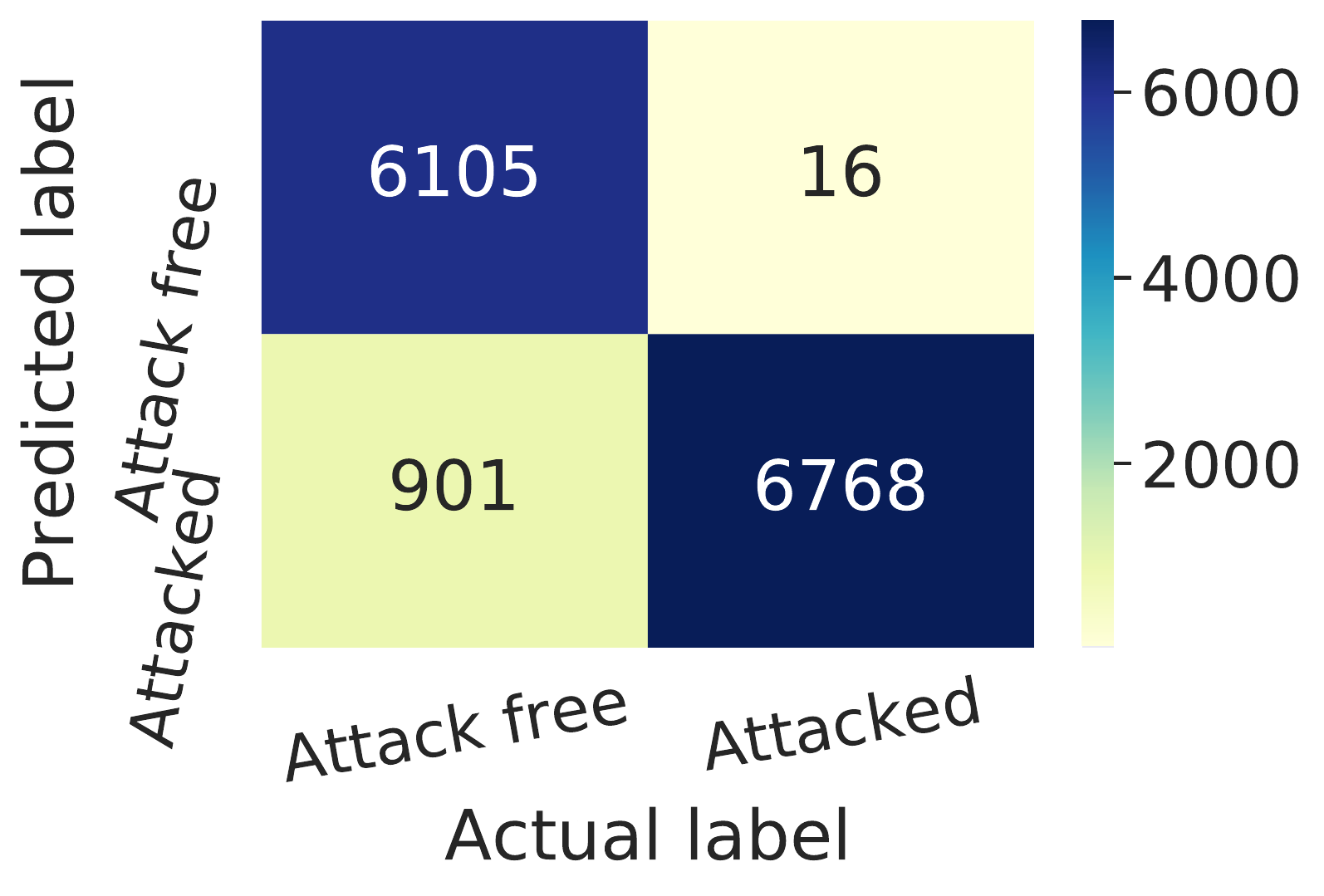}}
    \subfigure[]{\includegraphics[width=0.3\textwidth]{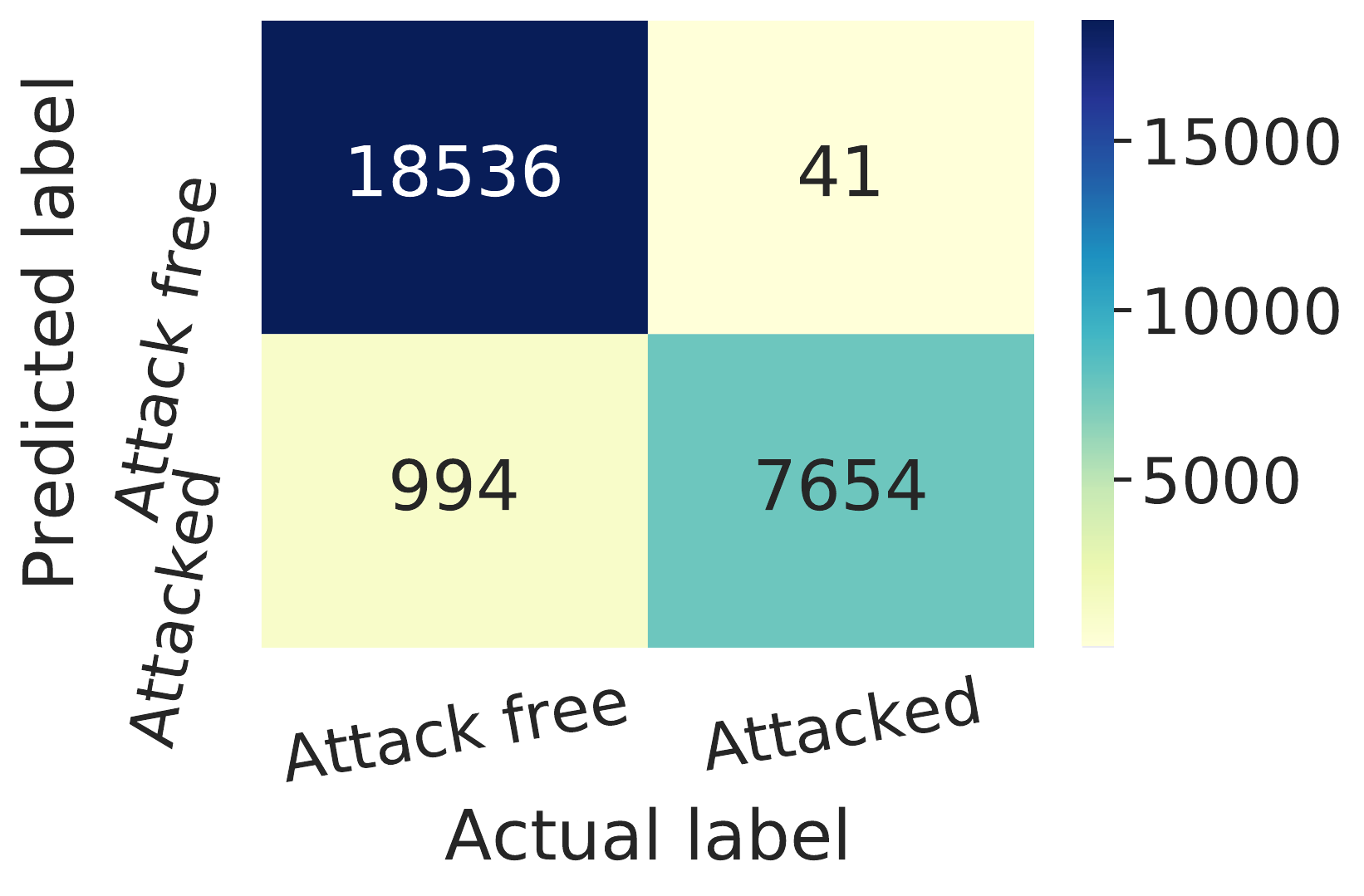}}
    \caption{The proposed methodology has a prediction accuracy of (a) 99.61\% considering DoS, (b) 99.83\% considering fuzzy, (c) 96.79\% considering spoofing, (d) 93.35\% considering replay, and (e) 96.20\% for the mixed attacks.}
    \label{fig:confusion_all}
    \vspace{0.60cm}
\end{figure*}

The proposed algorithm uses 67\% and 33\% data for training and testing, respectively. The proposed algorithm efficiently detected DoS, fuzzy, spoof, and replay attacked graphs and detect any mixed attacks. The proposed methodology has 99.61\%, 99.83\%, 96.79\%, and 93.35\% detection accuracy considering DoS, fuzzy, spoofing, and replay attacks, respectively. Figure~\ref{fig:confusion_all}(a), Figure~\ref{fig:confusion_all}(b), Figure~\ref{fig:confusion_all}(c), Figure~\ref{fig:confusion_all}(d) show the confusion matrix of DoS, fuzzy, spoofing, and replay attacks. One of the major advantages of using proposed techniques is it can detect mixed attacks. According to our analysis, the proposed methodology has 96.20\% detection accuracy for the mixed attacks, as shown in Figure~\ref{fig:confusion_all}(e).

Precision indicates the proportion of correctly predicted positive cases, and a high precision indicates the low false-positive rate of an algorithm~\cite{Powers:2020}. Recall tries to identify all actual positive cases, which is the ratio of correctly predicted positive observations to all the actual class observations. The F1 score is the harmonic mean of precision and recall. We analyze all three metrics for the DoS, fuzzy, spoofing, replay, and mixed attacks. We compare our method with a generic complement naive Bayes (CNB), multinomial naive Bayes (MNB), and the existing SVM-based~\cite{Tanksale:2019} and DCNN-bassed~\cite{Song:2020} IDSs. The proposed GNB-based methodology has excellent 99.79\% precision, which is better than all the competing methods. Using DoS attacks, the proposed method has 10.75\% and 38.77\% better recall and F1 scores, respectively, than the state-of-the-art SVM classifier. Overall, the proposed methodology has 99.11\%, 96.55\%, and 97.71\% of precisions, recall, and F1 scores, respectively.

\begin{table}[t!] \small 
\vspace{-0.70cm}
\renewcommand{\arraystretch}{1.5}
\captionsetup{aboveskip=-0.00cm,belowskip=0.5cm}
\caption{
		When considering DoS attacks, the proposed methodology has 53.79\%, 10.75\%, and 38.77\% better precision, recall, and F1 scores, respectively, than the state-of-the-art SVM classifier, using rawCAN data set~\cite{Lee:2017}.
		\label{tab:prtTable}}
		
\centering
\scalebox{0.55}{
\begin{tabular}{|c|c|c|c|c|c|c|c|c|c|c|c|c|c|c|c|}
\hline
\multirow{2}{*}{Types of attacks} & \multicolumn{3}{c|}{SVM~\cite{Tanksale:2019}}&
\multicolumn{3}{c|}{DCNN~\cite{Song:2020}}&
\multicolumn{3}{c|}{CNB} &\multicolumn{3}{c|}{MNB}  & \multicolumn{3}{c|}{GGNB} 
\tabularnewline

\cline{2-16}
& Pr & Re & F1& Pr & Re & F1 & Pr & Re & F1 & Pr & Re & F1 & Pr & Re & F1
\tabularnewline
\cline{1-16}

Dos &0.46&	0.89&0.61 & 1.0&0.9989&0.9995

&0.9969&1.0&	0.9985&	0.9966	&1.0&0.9983&0.9979&0.9975&	0.9977 
\tabularnewline
\cline{1-16}
Fuzzy &-&-&-&0.9995&0.9965&0.9980&

0.9945&1.0&0.9972&0.9899&1.0&0.9949&0.9989&0.9989&0.9989
\tabularnewline
\cline{1-16}
Spoofing &-&-&-&0.9999&0.9992&0.9995&

0.9599	& 0.9997&0.9794&0.9596&1.0&0.9794&0.9605&0.99634&	0.9781
\tabularnewline
\cline{1-16}

Replay &-&-&-& -&-&-&

0.9960	&0.8561&0.9208&0.9960&0.8576&0.9216&0.9974&0.8714&	0.9301
\tabularnewline
\cline{1-16}

Mixed &-&-&-&-&-&-&

0.8889	&0.9956&	0.9392&	0.8889&	0.9992&	0.9408&	0.9978	&0.9491&0.9728
\tabularnewline
\cline{1-16}

Overall &-&-&-&-&-&-& 

0.9498&0.9807&0.9650&0.9488&0.9822&0.9652&0.9911&0.9655&
0.9781
\tabularnewline
\cline{1-16}

\end{tabular}}
\vspace{-0.0cm}
\end{table}

\begin{figure*}
\captionsetup{aboveskip=-0.00cm,belowskip=-0.250cm}
    \centering
    \vspace{-0.5cm}
    \subfigure[]{\includegraphics[width=0.3\textwidth]{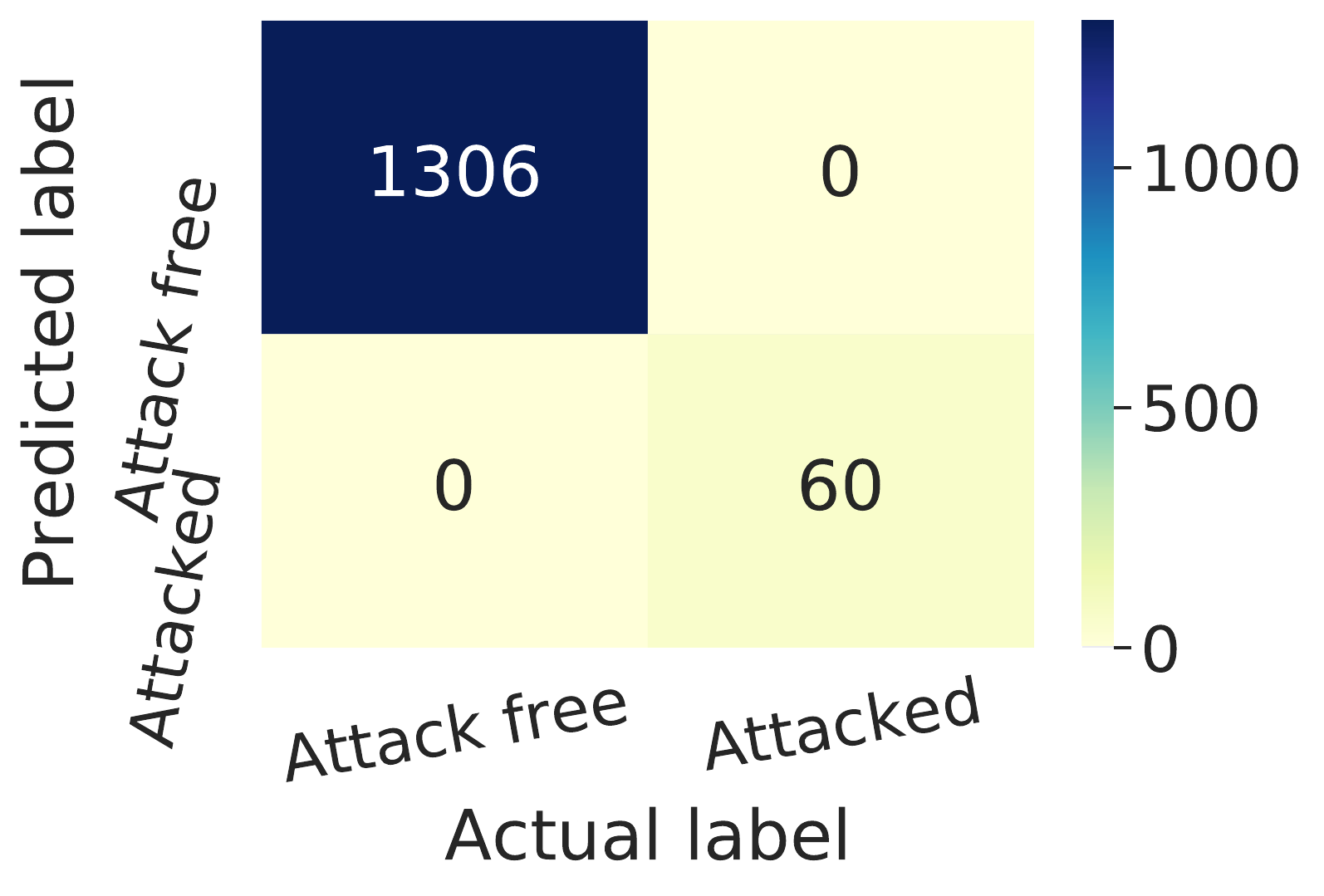}} 
     \subfigure[]{\includegraphics[width=0.3\textwidth]{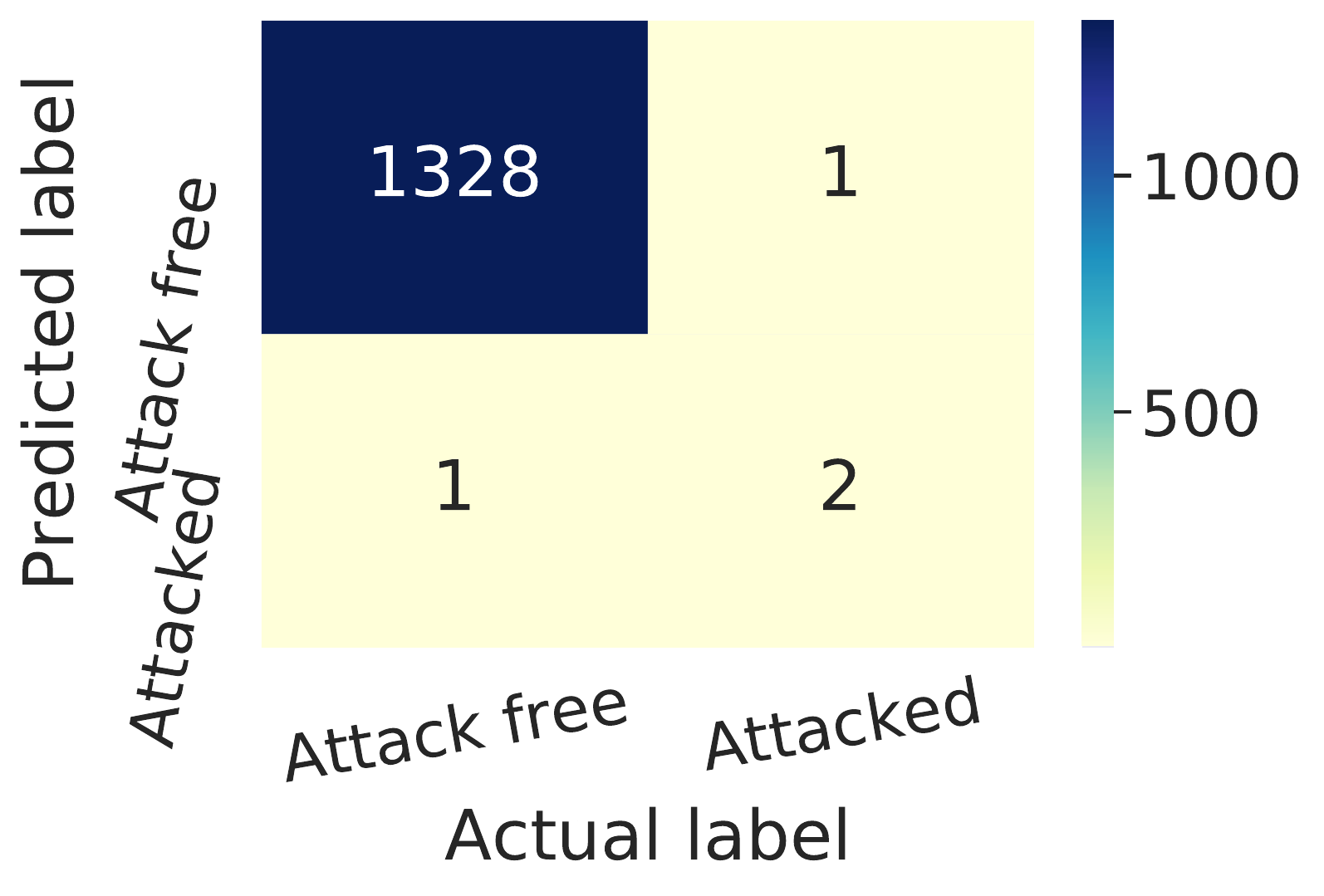}} 
    \subfigure[]{\includegraphics[width=0.3\textwidth]{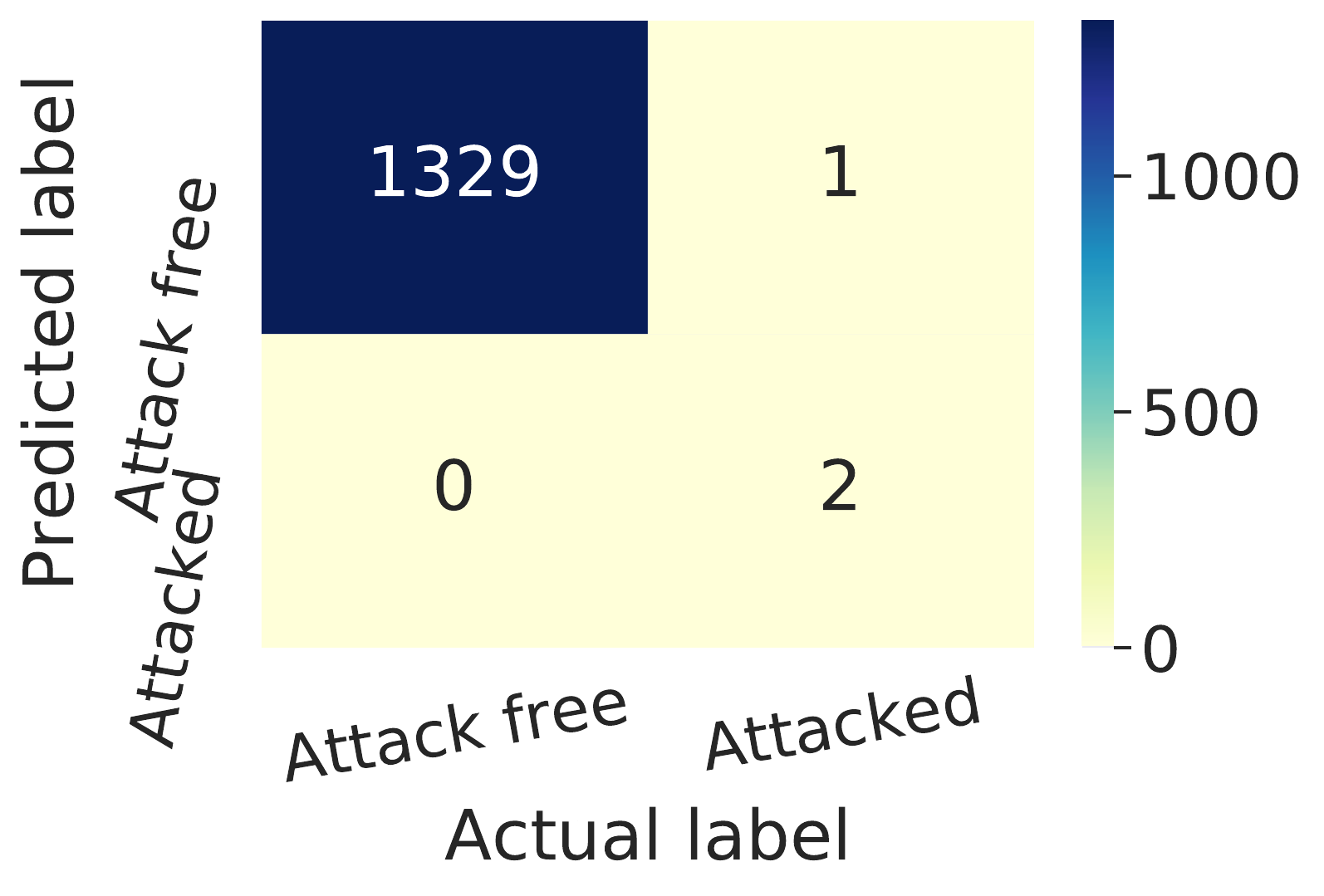}} 
    \subfigure[]{\includegraphics[width=0.3\textwidth]{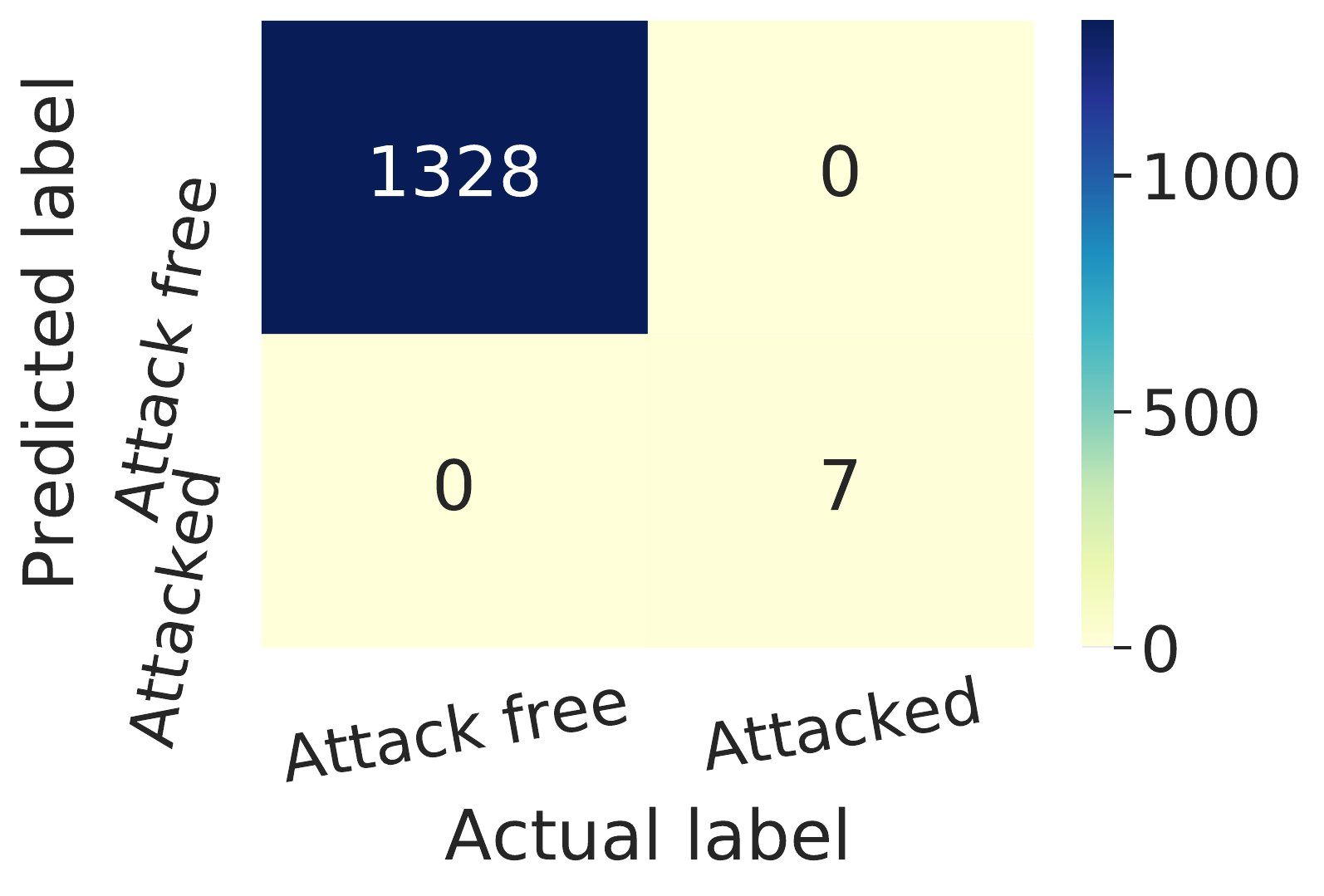}}
    \subfigure[]{\includegraphics[width=0.3\textwidth]{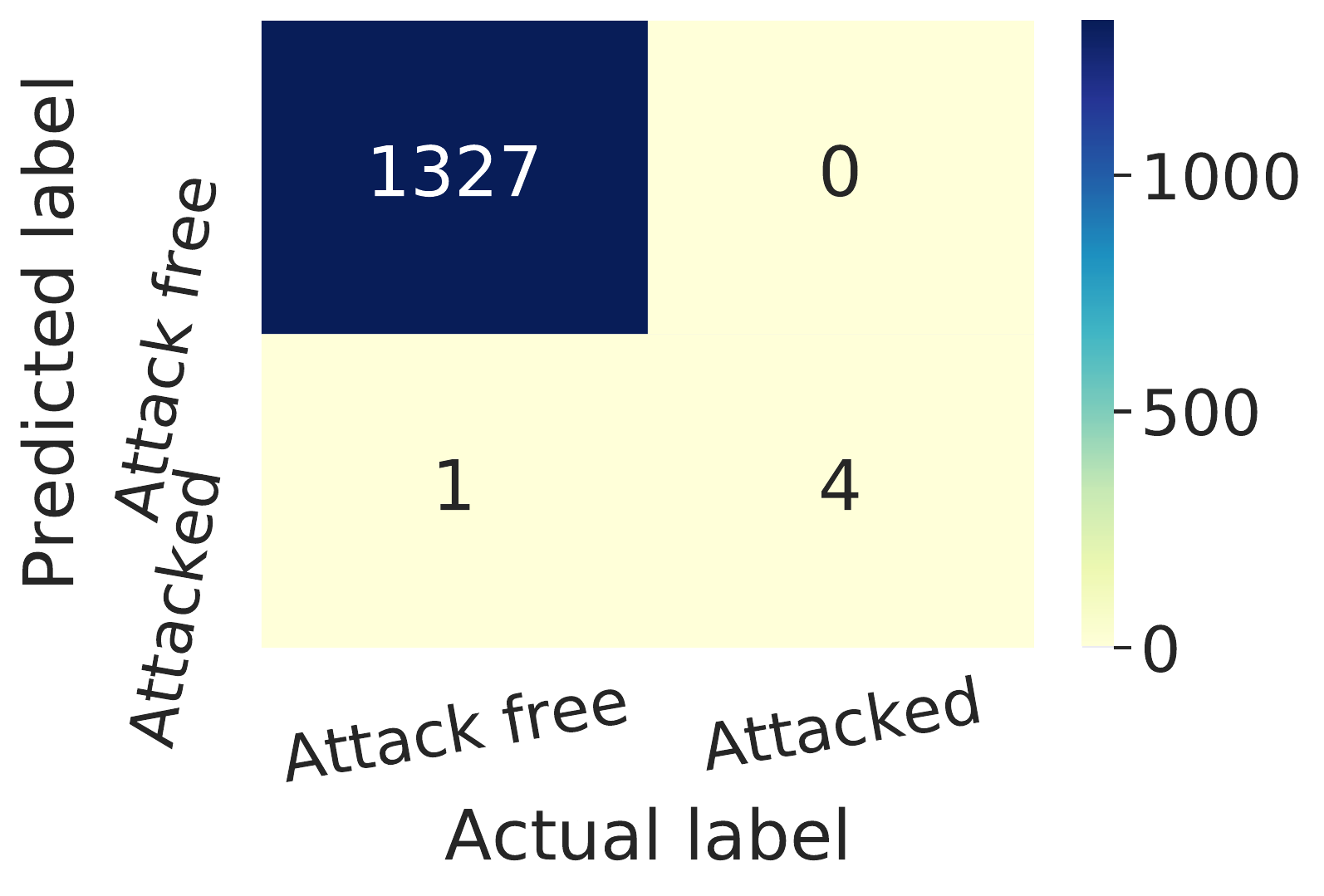}}
    \caption{The proposed methodology has excellent 100\%, 99.85\%, 99.92\%, 100\%, and 99.92\% detection accuracy considering DoS, diagnostic, fuzzing CAN ID, fuzzing payload, and replay, respectively.}
    \label{fig:confusion_opel_5}
    \vspace{0.50cm}
\end{figure*}

\begin{figure}
\captionsetup{aboveskip=-0.00cm,belowskip=-0.050cm}
    \centering
    \vspace{-0.05cm}
    \subfigure[]{\includegraphics[width=0.48\textwidth]{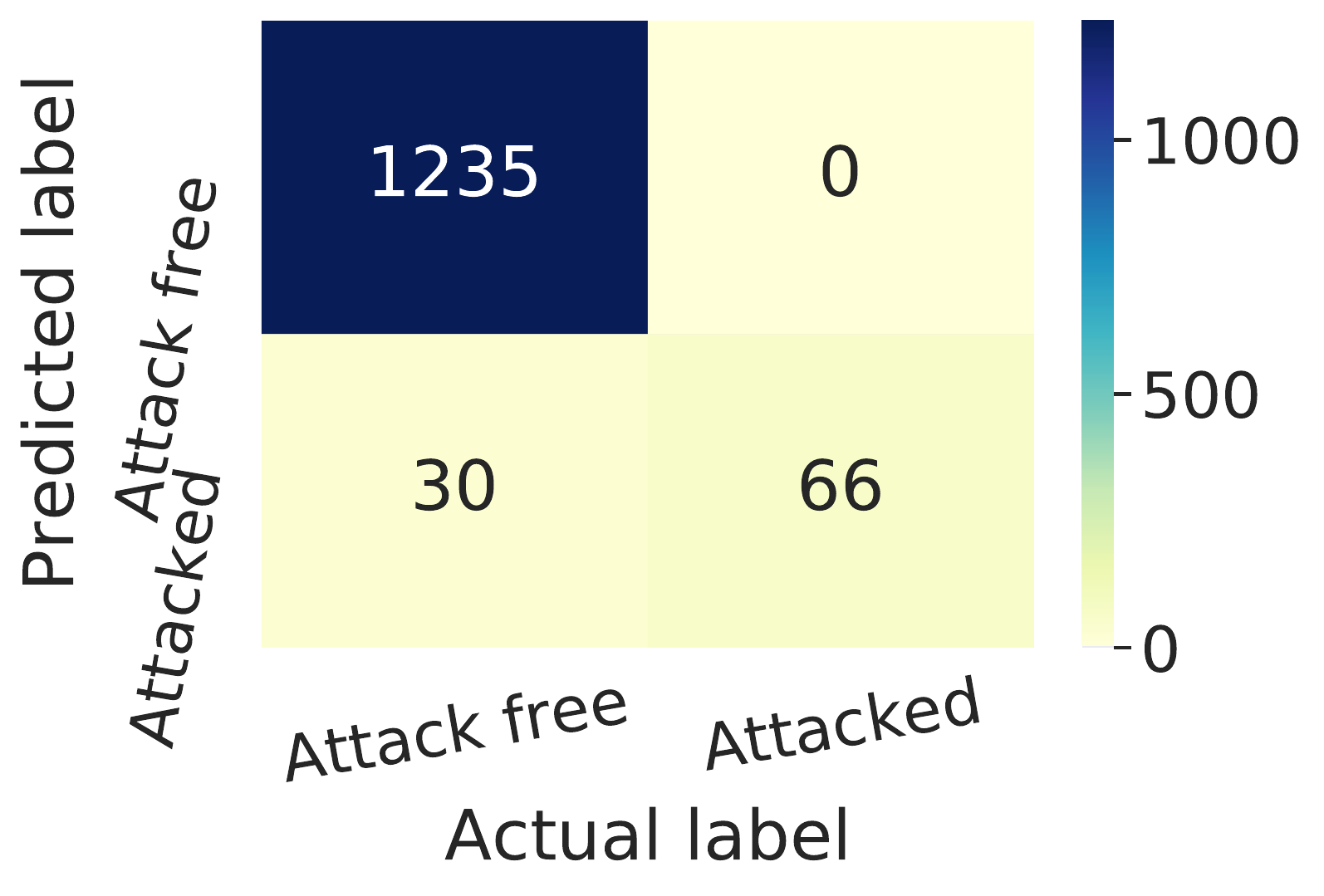}} 
    \subfigure[]{\includegraphics[width=0.48\textwidth]{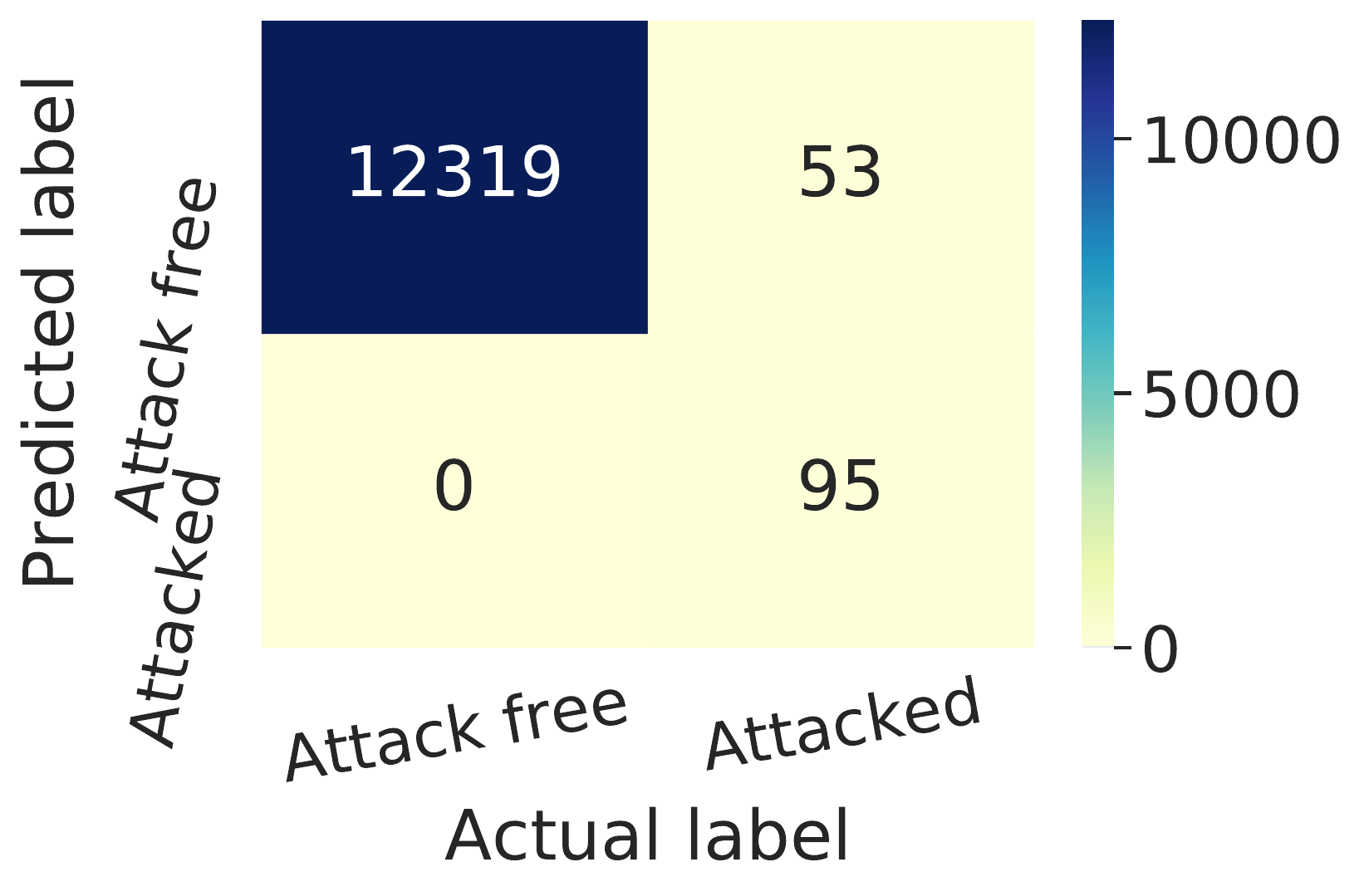}} 
    \caption{(a) The proposed methodology has 97.75\% detection accuracy considering suspension attacks, and (b) 99.57\% detection accuracy considering mixed attacks, which includes all six types of CAN attacks.}
    \label{fig:confusion_opel_sus_mixed}
    \vspace{0.05cm}
\end{figure}

In order to verify the efficiency of the proposed methodology, we used raw CAN attacked and attack free OpelAstra data set~\cite{Guillaume:2019}. The attacked data includes DoS, diagnostic, fuzzing CAN ID, fuzzing payload, replay, and suspension attacks. Besides, we consider the mixed attacks that have all six kinds of attacks. From the raw CAN bus data, we build $\sim 13.5K$, $\sim 17.6K$, $\sim 17.5K$, $\sim 17.5K$, $\sim 17.5K$, $\sim 17.5K$, and $\sim 17.5K$, and $\sim 37.7K$ attack free, DoS, diagnostic, fuzzing CAN ID, fuzzing payload, replay, suspension, and mixed attacked graphs. 
The proposed algorithm efficiently detected DoS, diagnostic, fuzzing CAN ID, fuzzing payload, replay, suspension, and detect any mixed attacks. The proposed methodology has 100\%, 99.85\%, 99.92\%, 100\%, and 99.92\% detection accuracy considering DoS, diagnostic, fuzzing CAN ID, fuzzing payload, and replay, respectively. 
Figure~\ref{fig:confusion_opel_5}(a), Figure~\ref{fig:confusion_opel_5}(b), Figure~\ref{fig:confusion_opel_5}(c), Figure~\ref{fig:confusion_opel_5}(d), and Figure~\ref{fig:confusion_opel_5}(e) show the confusion matrix of DoS, diagnostic, fuzzing CAN ID, fuzzing payload, replay attacks. 
In addition, the proposed methodology has 97.75\% and 99.57\% detection accuracy for the suspension and mixed attacks, as shown in Figure~\ref{fig:confusion_opel_sus_mixed}(a) and Figure~\ref{fig:confusion_opel_sus_mixed}(b), resspectively.

Similar to the other data set~\cite{Lee:2017}, we analyze precision, recall, and F1 scores for the DoS, diagnostic, fuzzing CAN ID, fuzzing payload, replay, suspension, and mixed attacks. Table~\ref{tab:prt2Table} shows the results of this analysis. The proposed GNB-based methodology has excellent 100\% precision, which is better than or comparable to all the competing methods. Using DoS attacks, the proposed method has 11\% and 39\% better recall and F1 scores, respectively, than the state-of-the-art SVM classifier. Overall, the proposed GGNB methodology has 99.84\%, 99.73\%, and 99.78\% of precision, recall, and F1 scores, respectively.

\begin{table*}[t!] \small 
\vspace{-0.00cm}
\renewcommand{\arraystretch}{1.5}
\captionsetup{aboveskip=-0.00cm,belowskip=-0.25cm}
\caption{
		When considering DoS attacks, the proposed methodology applying on the OpelAstra data set~\cite{Guillaume:2019} has 54\%, 11\%, and 39\% better precision, recall, and F1 scores, respectively, than the state-of-the-art SVM classifier.
		\label{tab:prt2Table}}
\centering 

\scalebox{0.57}{
\begin{tabular}{|c|c|c|c|c|c|c|c|c|c|c|c|c|c|c|r|}
\hline

\multirow{2}{*}{Types of attacks} & \multicolumn{3}{c|}{SVM~\cite{Tanksale:2019}}&
\multicolumn{3}{c|}{DCN~\cite{Song:2020}}&
\multicolumn{3}{c|}{CNB}&
\multicolumn{3}{c|}{MNB}& 
\multicolumn{3}{c|}{GGNB} 
\tabularnewline

\cline{2-16}
& Pr & Re & F1& Pr & Re & F1 & Pr & Re & F1 & Pr & Re & F1 & Pr & Re & F1
\tabularnewline
\cline{1-16}

DoS &0.46&	0.89&0.61 & 1.0&0.9989&0.9995 &
    1.0&1.0&1.0&
    1.0 &1.0& 1.0&
    1.0 &1.0 &1.0
\tabularnewline
\cline{1-16}
Diagnostic &-&-&-&-&-&-&
0.9932& 0.9985 &0.9959&
1.0 & 0.9977 & 0.9989&
0.9992& 0.9992& 0.9992
\tabularnewline
\cline{1-16}
Fuzzing canid &-&-&-&-&-&-&
1.0&1.0&1.0&
1.0& 1.0& 1.0&
1.0& 0.9992&0.9996
\tabularnewline
\cline{1-16}

Fuzzing payload &-&-&-&-&-&-&
1.0& 1.0 &1.0&
1.0& 1.0 &1.0&
1.0 &1.0 & 1.0
\tabularnewline
\cline{1-16}

Replay &-&-&-&-&-&-&

1.0& 1.0& 1.0&
1.0 & 0.9977 &0.9989&
0.9992& 1.0& 0.9996
\tabularnewline
\cline{1-16}

Suspension &-&-&-&-&-&-&
0.9960& 0.9992&0.9976&
1.0 & 0.9620& 0.9806&
0.9763 & 1.0& 0.9880
\tabularnewline
\cline{1-16}

Mixed  &-&-&-&-&-&-&
1.0 & 0.9972& 0.9986&
1.0 & 0.9940&0.9970&
1.0 & 0.9957& 0.9979

\tabularnewline
\cline{1-16}

Overall  &-&-&-&-&-&-&

0.9993& 0.9982& 0.9987&
1.0 &0.9936 & 0.9968&
0.9984& 0.9973 & 0.9978
\tabularnewline
\cline{1-16}

\end{tabular}}
\vspace{-0.6cm}
\end{table*}

\subsection{Sensitivity Analysis}
\label{subsec:sensitivity}
\begin{figure}[t]
	\begin{center}
		\vspace{-0.0cm} 
		\includegraphics[width = 0.6\textwidth]{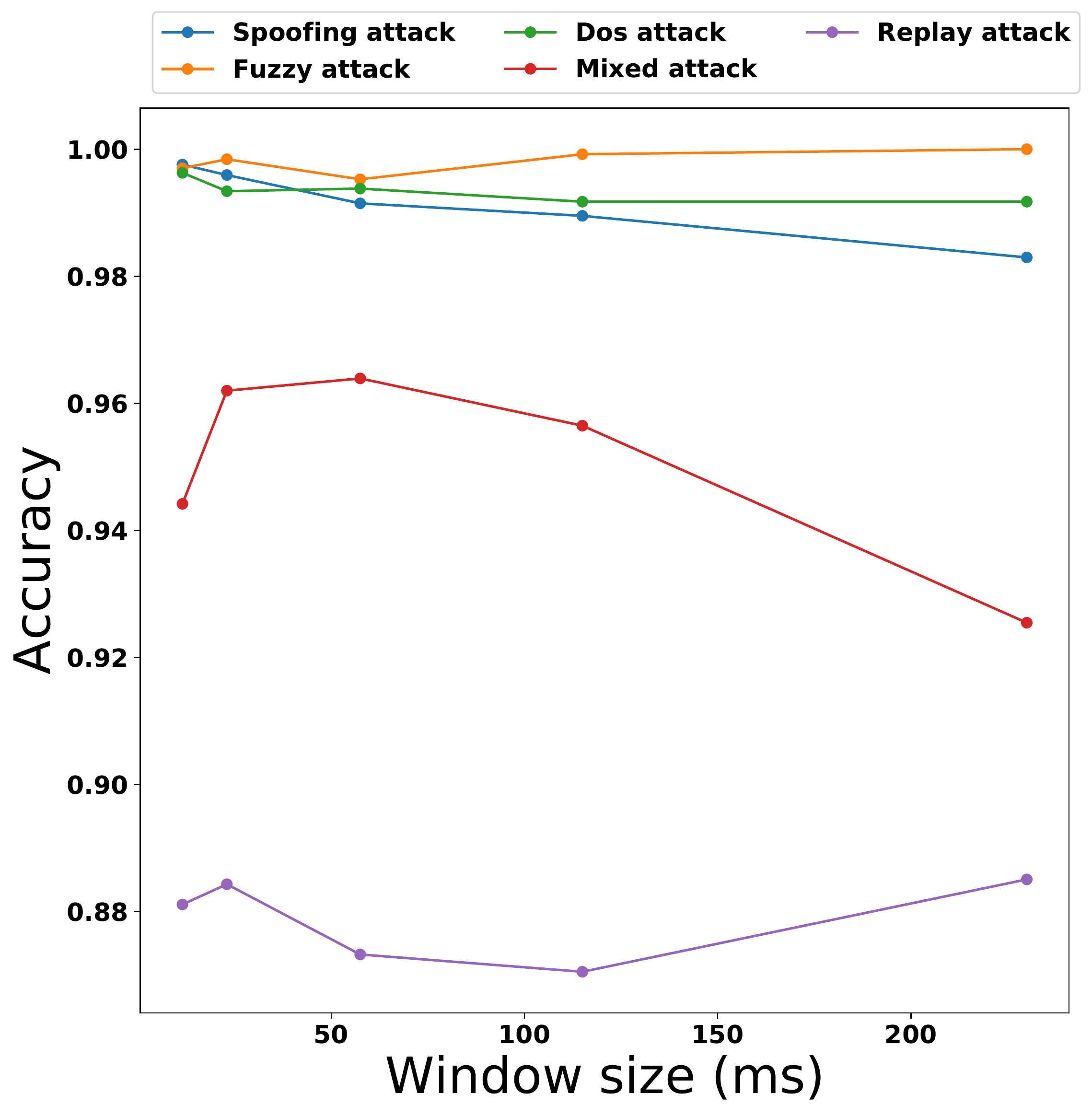}
		\vspace{-0.4cm} 
		\caption {The prediction accuracy varies by less than 1\% for DoS, fuzzy, spoofing, and replay attacks for 11.5ms to 230ms window size variation, and it exhibits about 2\% accuracy variation for the mixed attacks in the same window variation range using rawCAN~\cite{Lee:2017} data set.}
			\label{fig:sensitivity}
	\end{center}
	\vspace{-0.6cm}
\end{figure}

In order to measure the sensitivity of the proposed methodology for window size variation, we considered CAN communication window size from 11.5ms to 230ms. A wide range of window variation allows different CAN communication frequency. According to our analysis, the proposed method's accuracy is fairly consistent with the variation of window size, as shown in Figure~\ref{fig:sensitivity}. The proposed method has a less than 1\% accuracy variation in this extended window size variation considering DoS, spoofing, fuzzy, and replay attacks. The proposed method's accuracy dropped only from 94.42\% to 92.54\% for 11.5ms to 230ms window range for mixed attack. 

\subsection{Comparison of Time and Space Complexity}
\label{subsec:runtime_space}
For the GGNB algorithm, we compute the priors and the likelihood probabilities for each class in the training phase~\cite{Alpaydin:2020}. We compute the $n$ records prior in linear time $\mathcal{O}(n)$. Considering k features, we calculate k likelihoods for each class. The likelihood probabilities will be $\mathcal{O}(nk)$ for training two classes. For testing $m$ cases, the proposed GNB algorithm time complexity is $\mathcal{O}(mk)$. For training, the space complexity only considers the likelihoods for two classes of GNB, resulting in $\mathcal{O}(k)$. We computed the likelihoods during the training phase, so the training phase only requires a loop up, resulting in testing space complexity of $\mathcal{O}(m)$.

Generally, for a dataset with n records and $k$ features, the worst-case training time is $\mathcal{O}(kn^3)$. For $d_s$ support vectors, the test time is in the order of $\mathcal{O}(kd_s)$. For training, the space complexity only needs to store $d_s$ resulting in $\mathcal{O}(d_s)$. As expected, the proposed GGNB-based methodology has about $239\times$ and $135\times$ lower training and test time, respectively, compared to the existing SVM-based anomaly detection system.

\subsection{Features Reduction}
\label{subsec:feature_reduction}

\subsubsection{Correlation Matrix}
\label{subsubsec:correlation_matrix}

To correlate the features with the label of the data, we used correlation matrix~\cite{Bishop:2006}. The correlation coefficient can be expressed as
\begin{equation}
  r =
  \frac{ \sum_{i=1}^{n}(x_i-\bar{x})(y_i-\bar{y}) }{%
        \sqrt{\sum_{i=1}^{n}(x_i-\bar{x})^2}\sqrt{\sum_{i=1}^{n}(y_i-\bar{y})^2}}
\end{equation}
where r value in the range of -1 to +1. If two components x and y of the data are perfectly
correlated, then r = +1 or -1, and r = 0 if they are uncorrelated. The positive value of r indicates the proportional correlation of the components. On the other hand, the negative value of r indicates the inverse proportional correlation of the components.

\begin{figure}[t]
	\begin{center}
		\vspace{-0.15cm} 
		\includegraphics[width = 0.75\textwidth]{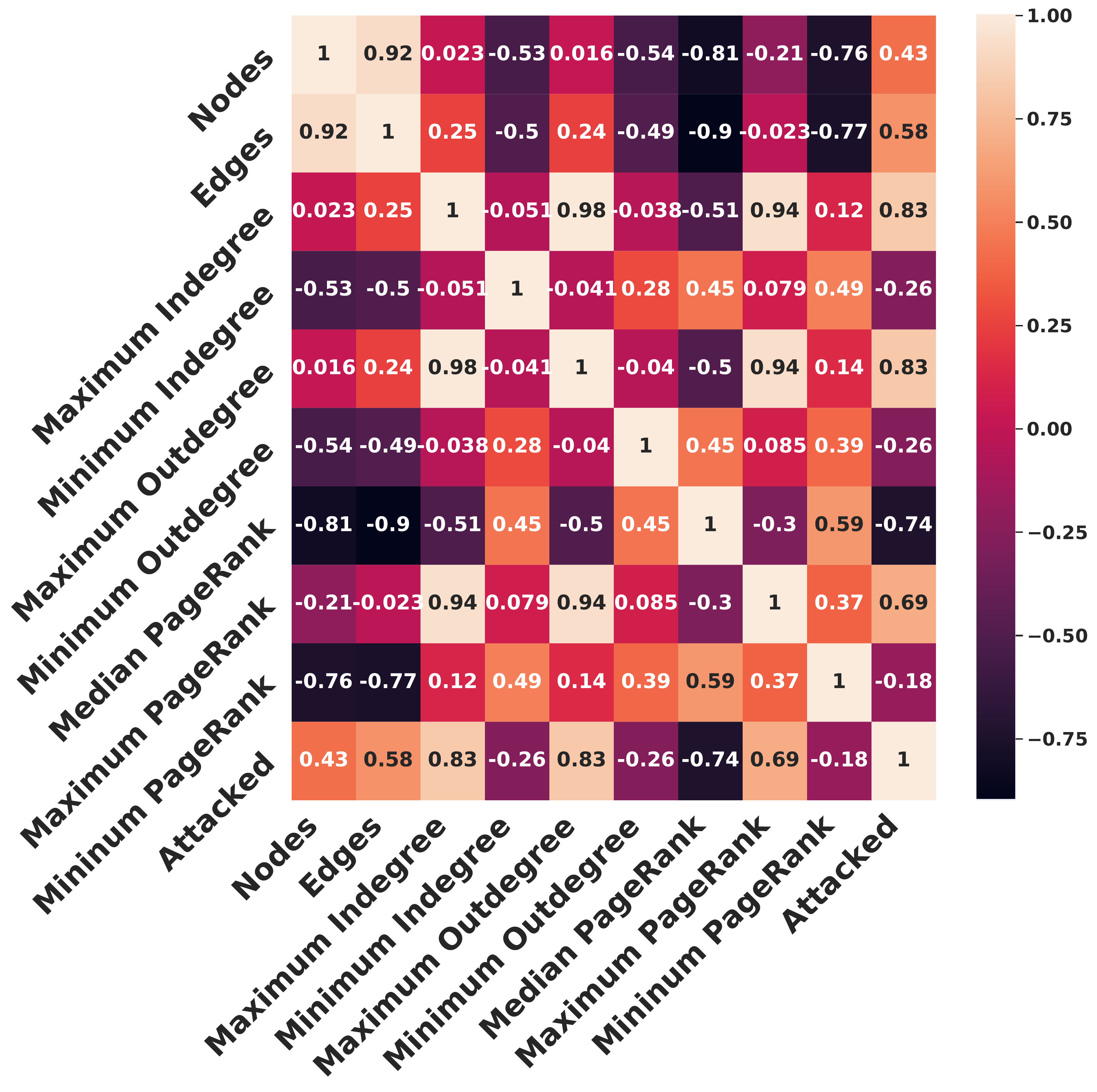}
		\vspace{-0.5cm} 
		\caption {The Maximum Indegree and Maximum Outdegree have the highest correlation with our desired output label or attack detection.}
			\label{fig:corr_matrix}
	\end{center}
	\vspace{-0.5cm}
\end{figure}

Feature reduction is the process of reducing the number of features in computation without losing important information and prediction accuracy \cite{Deepai:2020}. 
To reduce number of feature variables, we used correlation matrix for our proposed features, as shown in Figure~\ref{fig:corr_matrix}. Maximum Indegree and Maximum Outdegree are highly correlated with our desired output label (i.e., Attacked). Using only these two features the proposed algorithm exhibits 97.25 \% accuracy, using mixed attack data. When we incorporate two more features Median PageRank and Maximum PageRank for detecting attacked label, we achieved up to 98.06 \% accuracy.

\subsubsection{Feature Reduction Impact on Performance}
\label{subsubsec:performance_impact}
Feature reduction leads to the use of fewer resources to complete computations due to the decrease in the number of variables~\cite{Liu:2007}. Expectedly, feature reduction has a tremendous impact on the proposed methodology's performance, as shown in the Table~\ref{tab:relativetimeTable}. All the results are normalized with the GGNB algorithm with all nine features. For both the full set of features and the reduced number of features, the GGNB methodology has the lowest training time than other naive Bayes algorithms. On the other hand, CNB has the most down testing time than the other naive Bayes algorithms. With reduced features, the proposed GGNB algorithm has 18\% and 21\% lower training time using rawCAN~\cite{Lee:2017} and OpelAstra~\cite{Guillaume:2019} data sets, respectively. The GGNB has 37.5\% lower testing time using only four features on both the data sets with only a minor 0.03\% maximum accuracy penalty, as shown in Table~\ref{tab:relativetimeTable}.

\begin{table*}[t!] \small 
\vspace{-0.00cm}
\renewcommand{\arraystretch}{1.5}
\caption{With the reduced number of features, the proposed GGNB algorithm has 18\% and 21\% lower training time using rawCAN~\cite{Lee:2017} and OpelAstra~\cite{Guillaume:2019} data sets, respectively.
		\label{tab:relativetimeTable}}
\centering 
\scalebox{0.55}{
\begin{tabular}{|c|c|c|c|c|c|c|c|c|c|c|c|c|c|c|r|}
\hline

\multirow{2}{*}{Data set type} & 
\multirow{2}{*}{\# of features}&
\multicolumn{3}{c|}{CNB}&
\multicolumn{3}{c|}{MNB}& 
\multicolumn{3}{c|}{GGNB} 
\tabularnewline
\cline{3-11}
 & & Training time & Test time&Accuracy  & Training time & Test time  &Accuracy &Training time & Test time & Accuracy

\tabularnewline
\cline{1-11}
rawCAN~\cite{Lee:2017}& 9 &	1.1727 &	0.6337 &0.9075  &	1.1338   &	0.6849 &   0.9098   &	1.00  &	1.00 & 0.9620

\tabularnewline
\cline{2-11}
s
& 4 &	1.0555 &	0.5012 & 0.9647 & 1.0494	&	0.5733	&.  0.7174	  & 0.8196   &  0.6474    & 0.9618
\tabularnewline
\cline{1-11}

OpelAstra~\cite{Guillaume:2019}&	9	&	0.9024	&	0.5417	& 0.9080	&   0.8780  &	0.5417  &	0.9090 &	1.00	& 1.00	& 0.9090

\tabularnewline
\cline{2-11}

&	4	&    0.8415   &     0.4167   &    0.9090    &   0.8171      &  0.4583  &.    0.9090     &      0.7927            &   0.6250  &    0.9087
\tabularnewline
\cline{1-11}

\end{tabular}}
\vspace{-0.1cm}
\end{table*}

\begin{figure}[b!]
	\begin{center}
		\vspace{-0.05cm} 
		\includegraphics[width = 0.85\textwidth]{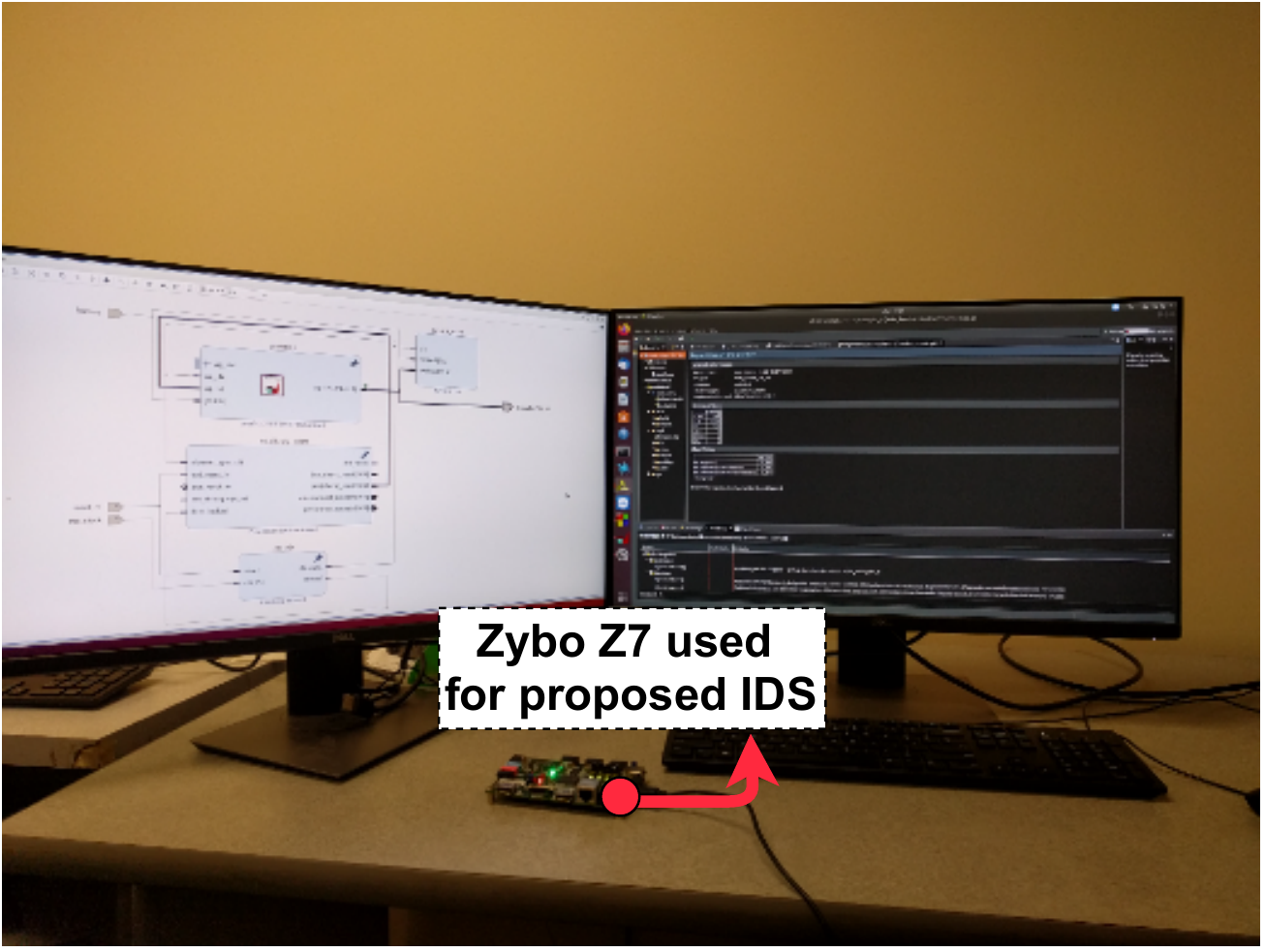}
		\vspace{-0.1cm} 
		\caption {The proposed GGNB requires $5.7 \times$, $5.9 \times$, $5.1 \times$, and $3.6 \times$ fewer slices, LUTs, flip-flops, and DSP units, respectively, than conventional NN architecture.}
			\label{fig:fpga_ids}
	\end{center}
	\vspace{-0.5cm}
\end{figure}

\subsubsection{FPGA Implementation Results}
\label{subsubsec:fpga}
We have implemented the proposed naive Bayes-based algorithms in a Xilinx Zybo Z7 board as shown in Figure~\ref{fig:fpga_ids}. Among different naive Bayes-based algorithms, 
CNB requires lower computation resources than the other two methodologies for attack prediction. For example, the proposed CNB requires $5.3 \times$, $4.7 \times$, $6.4\times$, and $6.5\times$ fewer slices, LUTs, flip-flops, and DSP cores, respectively, than the GGNB method. Moreover, the proposed GGNB requires $5.7 \times$, $5.9 \times$, $5.1 \times$, and $3.6 \times$ fewer slices, LUTs, flip-flops, and DSP units, respectively, compared to a conventional NN architecture with 4-hidden layers where each layer contains 500 neurons~\cite{Zhou:2019}.

\section{Conclusion}
\label{sec:conclusion}

We have presented the first-ever GGNB algorithm by using common graph properties and PR-related features. When we apply the
proposed algorithm on real rawCAN data set~\cite{Lee:2017}, the GGNB method has 99.61\%, 99.83\%, 96.79\%, and 93.35\% detection accuracy considering DoS, fuzzy, spoofing, and replay attacks, respectively. Moreover, this is the first-ever method with 96.20\% detection accuracy for mixed attacks. 

Using  OpelAsta data set~\cite{Guillaume:2019}, the proposed methodology has 100\%, 99.85\%, 99.92\%, 100\%, 99.92\%, 97.75\% and 99.57\% detection accuracy considering DoS, diagnostic, fuzzing CAN ID, fuzzing payload, replay, suspension, and mixed attacks, respectively.  

Better yet, the proposed GGNB-based methodology has about $239\times$ and $135\times$ lower training and test time, respectively, compared to the existing SVM classifiers.
Furthermore, the proposed correlation matrix-based feature reduction method reduces training time by 18\% and 21\% using rawCAN~\cite{Lee:2017} and OpelAstra~\cite{Guillaume:2019} data sets, respectively. Using a reduced number of features, the proposed method shows 37.5\% lower testing time on both the data sets with only a minor 0.03\% maximum accuracy penalty.

\clearpage


\bibliographystyle{elsarticle-num}
\bibliography{main}

\begin{thebibliography}{10}
\bibitem{Staggs:2013}
J.~Staggs, How to hack your mini cooper: reverse engineering {CAN} messages on
  passenger automobiles, Institute for Information Security.
\end{thebibliography}

\end{document}